\documentclass[titlepage]{article}

\usepackage{authblk}
\usepackage[T1]{fontenc}
\usepackage{ae}
\usepackage{aecompl}
\usepackage[utf8]{inputenc}
\usepackage[ngerman,english]{babel}
\usepackage{natbib}
\usepackage[export]{adjustbox} % Used to constrain images to a maximum size 
\usepackage{color} % Allow colors to be definedhttps://www.overleaf.com/9126683284mqrxtqnhwkpk
\usepackage{enumerate} % Needed for markdown enumerations to work
\usepackage[left=1in,right=1in]{geometry} % custom margins
\usepackage{longtable} % table across multiple pages
\usepackage{arydshln} % dashed lines within tables
\usepackage{rotating} % for landscape table
\usepackage{pdflscape} % another idea for landscape table, rotating entire page within pdf
\usepackage{amsmath}
\usepackage{mathtools}
\usepackage{amssymb}
\usepackage{pbox} % for making line breaks within table cells
\usepackage{booktabs}
\usepackage{fancyvrb} % verbatim replacement that allows latex
\usepackage{grffile} % extends the file name processing of package graphics 
\usepackage{pifont}% http://ctan.org/pkg/pifont

\usepackage{color}
\usepackage{todonotes}
\usepackage{float}
\usepackage{wrapfig}
\usepackage{subcaption}
\usepackage{setspace} % for double spacing
\usepackage{multirow} % merged cells in table
\usepackage{hyperref} % different colors for citations and links
\hypersetup{
      breaklinks=true,  % so long urls are correctly broken across lines
      colorlinks=true,
      urlcolor=blue,
      linkcolor=blue,
      citecolor=blue,
}
\usepackage{fancyhdr}
\usepackage{abstract}

\usepackage{tikz}
\usetikzlibrary{positioning}% To get more advances positioning options
\tikzset{every picture/.style={font=\sffamily}}

\usepackage{lineno}
\modulolinenumbers[5]
\definecolor{darkgreen}{rgb}{0.05,0.35,0.1}
\newcommand{\revised}[1]{#1}
\newcounter{result}[section]

\title{\rule{\textwidth}{1pt}\\\LARGE{\revised{\textbf{
Machine-Learning-Based Diagnostics \\of EEG Pathology
}}}\\\rule{\textwidth}{1pt}\\\
\\
\normalsize{Keywords: \itshape{
Machine Learning, 
Electroencephalography, 
Diagnostics, 
Pathology, \\
Features, 
Riemannian geometry, 
Convolutional Neural Networks
}}
}

\author[a,b,c]{Lukas A. W. Gemein\thanks{Corresponding author: lukas.gemein@uniklinik-freiburg.de\\ \textit{Submitted preprint\hfill December 17, 2019}}}
\author[a,b]{Robin T. Schirrmeister}
\author[a,b]{Patryk Chrabąszcz}
\author[a]{Daniel Wilson}
\author[c]{Joschka Boedecker}
\author[d]{Andreas Schulze-Bonhage}
\author[b]{Frank Hutter}
\author[a, d]{Tonio Ball}
\affil[a]{
    Neuromedical AI Lab,
    Department of Neurosurgery, 
    Medical Center – University of Freiburg,
    Faculty of Medicine, University of Freiburg, 
    Engelbergerstr. 21, 79106 Freiburg, Germany}
\affil[b]{
    Machine Learning Lab, 
    Computer Science Department – University of Freiburg,
    Faculty of Engineering, University of Freiburg, 
    Georges-Köhler-Allee 74, 79110 Freiburg, Germany}
\affil[c]{
    Neurorobotics Lab, 
    Computer Science Department – University of Freiburg,
    Faculty of Engineering, University of Freiburg, 
    Georges-Köhler-Allee 80, 79110 Freiburg, Germany}
\affil[d]{
    Freiburg Epilepsy Center,
    Department of Neurosurgery, 
    Medical Center – University of Freiburg,
    Faculty of Medicine, University of Freiburg, 
    Breisacher Str. 64, 79106 Freiburg, Germany}

\newcommand{\beginsupplement}{%
        \setcounter{table}{0}
        \renewcommand{\thetable}{S\arabic{table}}%
        \setcounter{figure}{0}
        \renewcommand{\thefigure}{S\arabic{figure}}%
     }
\begin{document}
\lfoot{\vfill© 2019. Licensed under the CC-BY-NC-ND license \url{creativecommons.org/licenses/by-nc-nd/4.0/}.}
\maketitle

%% Uncomment line below for double spacing
%\doublespacing

%\linenumbers

\begin{abstract}
Machine learning (ML) methods have the potential to automate clinical EEG analysis. They can be categorized into feature-based (with handcrafted features), and end-to-end approaches (with learned features). Previous studies on EEG pathology decoding have typically analyzed a limited number of features, decoders, or both. For a I) more elaborate feature-based EEG analysis, and II) in-depth comparisons of both approaches, here we first develop a comprehensive feature-based framework, and then compare this framework to state-of-the-art end-to-end methods. To this aim, we apply the proposed feature-based framework and deep neural networks including an EEG-optimized temporal convolutional network (TCN) to the task of pathological versus non-pathological EEG classification. For a robust comparison, we chose the Temple University Hospital (TUH) Abnormal EEG Corpus (v2.0.0), which contains  approximately 3000 EEG recordings. The results demonstrate that the proposed feature-based decoding framework can achieve accuracies on the same level as state-of-the-art deep neural networks. We find accuracies across both approaches in an astonishingly narrow range from 81--86\%. Moreover, visualizations and analyses indicated that both approaches used similar aspects of the data, e.g., delta and theta band power at temporal electrode locations. We argue that the accuracies of current binary EEG pathology decoders could saturate near 90\% due to the imperfect inter-rater agreement of the clinical labels, and that such decoders are already clinically useful, such as in areas where clinical EEG experts are rare. We make the proposed feature-based framework available open source and thus offer a new tool for  EEG machine learning research.
\end{abstract}

%\normalsize{Keywords: \itshape{
%Machine Learning, 
%Electroencephalography, 
%Diagnostics, 
%Pathology, \\
%Features, 
%Riemannian geometry, 
%%Deep Learning, 
%Convolutional Neural Networks
%%, Automated Machine Learning
%}}

\section{Introduction}
There is a great interest in using machine learning (ML) methods for automatic electroencephalogram (EEG) analysis, especially in the domain of clinical diagnostics based on the EEG. 
For example, ML has an important role in developing Brain-Computer Interfaces (BCIs) to support paralyzed people [\cite{schroer2015autonomous}] or to improve neurological rehabilitation [\cite{ramos2013brain}, \cite{tangermann2014towards}, \cite{van2015brain}]. 
It also forms a basis for detecting and predicting epileptic seizures [\cite{subasi2019seizure}, \cite{hugle2018early}, \cite{kiralkornek2018seizure}, \cite{mirowski2009classification}] with the goal of warning patients of upcoming seizures or to control brain stimulation for preventing or stopping seizure activity. 
Furthermore, ML allows for the automation of the process of EEG-based sleep staging [\cite{biswal2017sleepnet}] and neurological diagnostics of both specific diseases and disorders such as Alzheimer's
disease [\cite{lehmann2007application}], depression [\cite{cai2016pervasive}, \cite{hosseinifard2013classifying}], traumatic brain injuries [\cite{albert2016automatic}], and
strokes [\cite{giri2016ischemic}], or of general EEG pathology [\cite{abnormalLopez}, \cite{schirrmeister2017patho}, \cite{roy2019chrononet}, \cite{amin2019cognitive}, \cite{alhussein2019eeg}, \cite{van2019detecting}].

There are several facts that motivate the interest in automatic clinical EEG diagnosis. 
First, the evaluation of clinical EEGs is frequently a time-consuming and exhausting process. 
Second, it requires years of training to assess pathological changes in clinical EEG recordings. 
Moreover, even for highly trained EEG experts, diagnostic accuracy is subject to a number of limitations. 
It depends highly on individual training and experience, consistency of rating over time, time constraints in different filter settings of frequently subjectively defined frequency bands, and unclear criteria for the thresholding of potential changes, e.g., at low amplitude in relation to the background EEG. 
Accordingly, inter-rater agreement in assessing EEGs is known to be moderate [\cite{landis1977measurement}], i.e., \cite{grant2014eeg} found a Fleiss' Kappa of 0.44 when neurologists classified recordings to one of seven classes including seizure, slowing, and normal activity. 
In the more general task of classifying EEG recordings as pathological or normal, \cite{houfek1959reliability} and \cite{rose1973reliability} reported inter-rater agreements of 86\% and 88\% based on two neurologists.
The development of algorithms for automated EEG diagnostics could support clinicians in screening EEGs.
They could not only reduce the workload of clinicians, but also allow for earlier detection and treatment of diseases, which could enhance patient care. 
Furthermore, they could provide high-quality EEG interpretation and classification to patients that cannot attend specialized centers.

We broadly categorize ML for EEG analysis into two approaches: feature-based and end-to-end methods. 
Feature-based decoding methods have a long history of successful application in different EEG decoding tasks.
In this approach, typically handcrafted and \textit{a priori} selected features represent the data. 
For example, a researcher could \textit{a priori} decide to use the spectral power in certain frequency bands as features, if they assume that these bands are informative for the decoding task at hand. 
The choice of exact frequency bands could then be handcrafted, such as in the common spatial patterns (CSP) algorithm for motor decoding [\cite{muller1999designing}], or they could be determined by automatic feature selection, such as by the recursive band estimation in the filter bank CSP (FBCSP) algorithm [\cite{ang2008filter}].
This procedure relies on the domain expertise of the researcher. 
If the \textit{a priori} feature decisions are sub-optimal, it can diminish the quality of the resulting analysis. 
Conversely, owing to its explicit nature, interpretability of the classification decisions is frequently considered an advantage of feature-based decoding.

Conversely, end-to-end decoding methods accept raw or minimally preprocessed data as inputs. 
To date, end-to-end deep learning has attracted attention primarily owing to its success in other research fields, such as computer vision [\cite{krizhevsky2012imagenet}] and speech recognition [\cite{hinton2012deep}]. 
However, it has also recently gained momentum through the successful application of deep learning with artificial neural networks to EEG analysis [\cite{craik2019deep}]. 
By design, the networks learn features themselves and allow for a joint optimization of the feature extraction and classification. 
This procedure can lead to superior solutions or the discovery of unexpected informative features and does not require handcrafting, at least not for the extraction of the features. 
End-to-end models have the reputation for being ``black boxes'' with regard to the learned features; it is a challenge and an ongoing topic of intense research in the deep learning field to understand what they have learned [\cite{montavon2018methods}, \cite{sturm2016interpretable}, \cite{hartmann2018hierarchical}]. 
Another common concern is that the complexity in the application of ML is only shifted from the domain of feature engineering in traditional approaches to the domain of network engineering inasmuch as it could be necessary to handcraft the networks according to the requirements of a given task. 

In the literature, there is a lack of systematic comparisons of traditional feature-based versus end-to-end ML analysis of EEGs, despite their importance for a wide range of applications.
In particular, there are no studies comparing the accuracy of pathology decoding from EEGs using a broad range of time, frequency, and connectivity features with well-established end-to-end methods based on a large EEG data set for a robust comparison. 
Past comparisons of deep learning results have frequently only considered other deep learning results, or (rather) simple feature-based baselines (using thresholds, linear discriminant analysis, or linear regression) with limited feature sets.
This can lead to unfair comparisons among the methods. Moreover, it can create the impression of superiority of one approach over another. 
It is actually possible that deep learning could not yield an improvement over feature-based decoding in specific applications. 
Recently, \cite{rajkomar2018scalable} demonstrated that logistic regression can compete with deep neural networks in predicting medical events from electronic health records. 
To the best of our knowledge, there is no work that compares different deep neural network architectures with a feature-based approach, especially using a large set of features of several domains to decode the EEGs. 
However, we anticipate that large-scale comparisons between feature-based and end-to-end methods will be critical to the advancement of ML techniques for EEGs beyond the current state-of-the-art. 
Developing methods in both of these important fields in a mutually informed manner will likely be fruitful for both advanced feature-based and novel end-to-end EEG methodologies. 

In this paper, we compare end-to-end decoding using deep neural networks to feature-based decoding using a large set of features. 
We design a comprehensive study using the Temple University Hospital (TUH) Abnormal EEG Corpus [\cite{abnormalLopez}] with approximately 3000 recordings of at least 15 min duration each. 
This is a subset of the TUH EEG Corpus [\cite{obeid2016temple}], the largest publicly available collection of EEG recordings to date. 
For feature-based pathology decoding, we use random forest (RF) [\cite{breiman2001random}], support vector machine (SVM) [\cite{boser1992training}], Riemannian geometry (RG), and the auto-sklearn calssifier (ASC) [\cite{feurer2015efficient}] – an automated ML toolkit. 
For end-to-end pathology decoding, we use three types of convolutional neural networks (ConvNets, in other publications also CNN) [\cite{lecun1999object}] that have a history of successful application in different EEG decoding tasks. These are the 4-layer ConvNet architecture Braindecode Deep4 ConvNet (BD-Deep4), which has been successfully applied to motor decoding [\cite{schirrmeister2017hbm}], velocity and speed decoding [\cite{hammer2013role}], and pathology decoding [\cite{schirrmeister2017patho}, \cite{van2019detecting}]. 
Importantly, we use \href{https://github.com/TNTLFreiburg/braindecode}{Braindecode (BD)\footnote{\label{note:braindecode}available for download at \url{https://github.com/TNTLFreiburg/braindecode}}}, a previously developed and evaluated deep learning toolbox for EEGs, ``out of the box'' – without task-specific network engineering, i.e., without adaptation to the architectures.
Furthermore, we use a TCN [\cite{bai2018empirical}] that is optimized for EEG decoding with a neural architecture search.
We call this adaptation BD-TCN.

To the best of our knowledge, there are currently six published results for pathology decoding from EEGs, five of which used the TUH Abnormal EEG Corpus (Table~\ref{tab:related_works}).
However, only one publication [\cite{abnormalLopez}] used handcrafted features and a classification through a CNN with a multi layer perceptron (MLP).
All other papers considered this initial feature-based decoding result as a baseline. 
Whereas \cite{amin2019cognitive} and \cite{alhussein2019eeg} have reported the highest accuracies in decoding pathology from EEGs, we exclude them from our direct comparison. 
The papers mention ``pretrained models'' and additional ``10000 normal EEG recordings'', which would appear to be an extension of the TUH Abnormal EEG data set without specifying more details. 
In ML, the effect of more data is commonly greater than the effect of more elaborate algorithms [\cite{halevy2009unreasonable}]. 
A direct comparison of publications with access to a substantially larger amount of training data would hence be unfair.

\begin{table}
    \centering
    \begin{tabular}{c|c|c|c}
        \textbf{Automated EEG Diagnosis} & \textbf{Features} &\textbf{Architecture} & \textbf{ACC [\%]} \\ \hline
        \cite{abnormalLopez} & Cepstral coeff. & CNN + MLP & 78.8 \\
        \cite{schirrmeister2017patho} & & BD-Deep4 & 85.4 \\
        \cite{roy2019chrononet} & & ChronoNet & 86.6   \\
        \cite{amin2019cognitive}* & & AlexNet + SVM & 87.3 \\
        \cite{alhussein2019eeg}* & & 3 x AlexNet + MLP & 89.1 \\
        \cite{van2019detecting}$^+$ & & BD-Deep4 & 82.0
    \end{tabular}
    \caption{Related works on pathology decoding using TUH Abnormal EEG Corpus. All approaches rely on ConvNet architectures. Only chronologically oldest publication used handcrafted features. Publications marked with * used pretrained models and additional training data. Publication marked with $^+$ did not use TUH Abnormal EEG Corpus.}
    \label{tab:related_works}
\end{table}

The paper is structured as follows. 
In Section~\ref{sec:methods}, we provide an introduction to the TUH Abnormal EEG Corpus upon which we base our study. 
We then discuss the feature and deep learning pipeline in detail and explain how we proceeded in evaluating and comparing both approaches. 
The section closes with a discussion of the analytical methods we used to assist in our interpretation of the results. 
In Section~\ref{sec:results}, we present and discuss our results including an extensive comparison of both pipelines.
We present a general discussion in Section~\ref{sec:discussion_new}, and close with a brief outlook and conclusions in Section~\ref{sec:outlook}.

\section{Material and Methods}
\label{sec:methods}
\subsection{Data}
We base our study on the \href{https://www.isip.piconepress.com/projects/tuh\_eeg/html/downloads.shtml}{TUH Abnormal EEG Corpus}\footnote{available for download at \url{https://www.isip.piconepress.com/projects/tuh\_eeg/html/downloads.shtml}} (v2.0.0), which is currently unique owing to its size and public availability and has enabled the task of general pathology decoding from EEGs.
The corpus includes 2993 recordings of at least 15 min duration obtained from 2329 unique patients and consists of a development and separate final evaluation set (Table~\ref{tab:tuh_abnormal}).
It contains recordings of both male and female patients of a wide age range (7 days to 96 years), thus including infants, children, adolescents, adults, and senior patients. 
Pathologies diagnosed in the patients in the data set include (but are not limited to) epilepsy, strokes, depression, and Alzheimer's disease. 
The data set includes physician reports that provide additional information regarding each EEG recording, such as main EEG findings, ongoing  medication of the patient, and medical history. 
In the description of the data set\footnote{\url{https://www.isip.piconepress.com/projects/tuh\_eeg/downloads/tuh\_eeg\_abnormal/v2.0.0/\_AAREADME.txt}}, the TUH reports an inter-rater agreement of 97--100\%.
In the literature, the reported scores are typically considerably lower [\cite{houfek1959reliability}, \cite{rose1973reliability}].
The almost perfect rating scores could be a consequence of the review process of the findings that were performed by medical students that knew the diagnoses beforehand [\cite{picone2019agreement}].
For more information on the data set see \cite{abnormalLopez} and \cite{obeid2016temple}.

\begin{table}[ht!]
    \centering
    \begin{tabular}{c|cc|cc||c}
        \textbf{TUH Abnormal EEG} & \multicolumn{2}{c|}{\textbf{Non-pathological}} & \multicolumn{2}{c||}{\textbf{Pathological}} & \textbf{Intersection} \\
        \textbf{Corpus (v2.0.0)} & \textbf{Recordings} & \textbf{Patients} & \textbf{Recordings} & \textbf{Patients} & \textbf{Patients} \\ \hline
        \textbf{Development set} & 1371 & 1237 & 1346 & 893 & 54 \\
        \textbf{Final evaluation set} & 150 & 148 & 126 & 105 & 0 \\ \hline
        \textbf{Total} & 1521 & 1385 & 1472 & 998 & 54
    \end{tabular}
    \caption{Number of recordings and patients in TUH Abnormal EEG Corpus (v2.0.0). For certain patients, there exist several recordings. For other patients in the development set, there exist normal and abnormal recordings. There is no overlap of patients in development and the final evaluation set.}
    \label{tab:tuh_abnormal}
\end{table}

\subsection{Common preprocessing in both feature and end-to-end pipeline} 
\label{sec:preprocessing}
\begin{figure}[ht!]
  \begin{center}
    \includegraphics[width=0.48\textwidth]{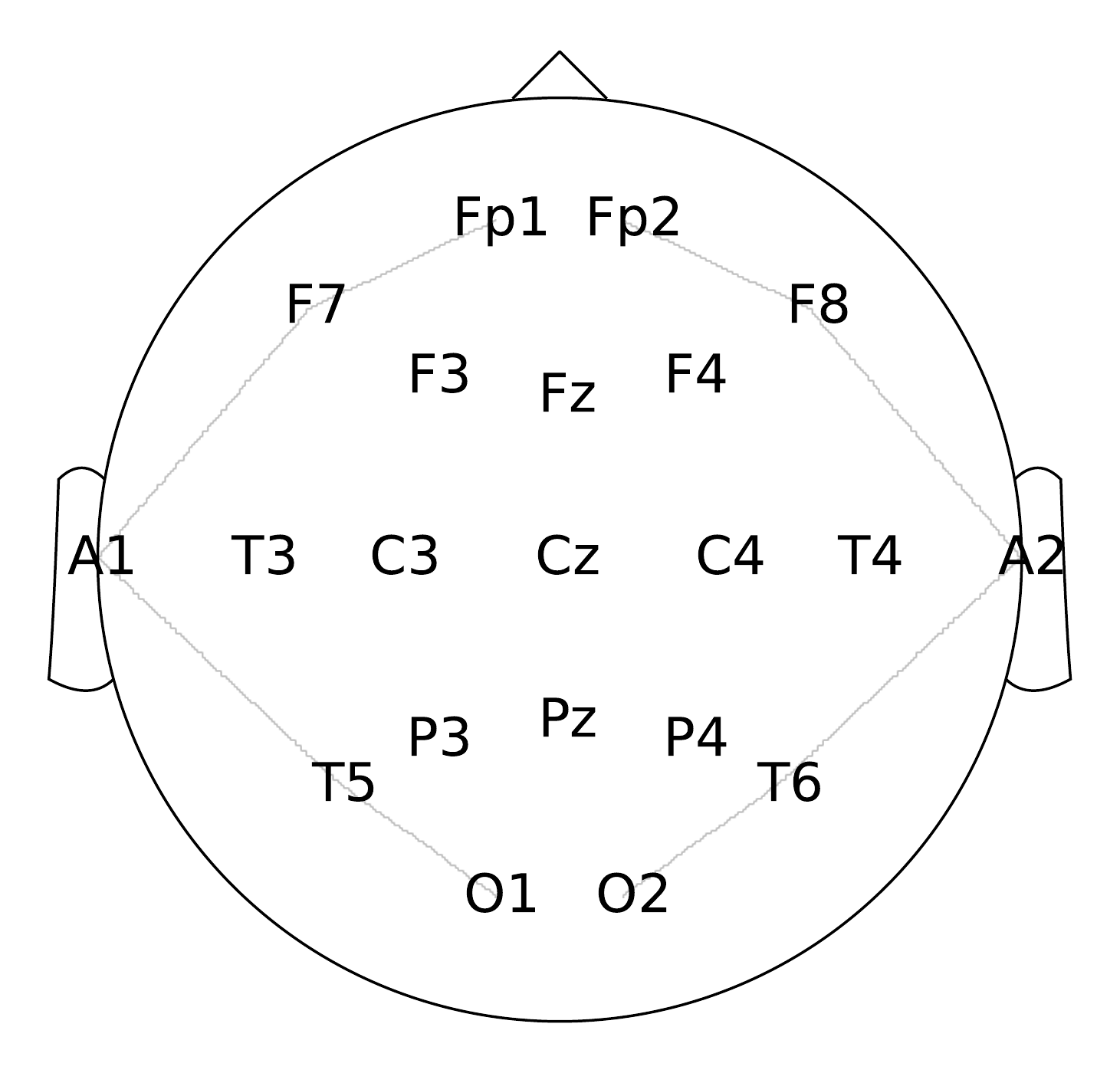}
  \end{center}
    \caption{Topographical map of 21 EEG channel subset of international 10-20 placement [\cite{jasper1958report}] common in all recordings included in TUH Abnormal EEG Corpus (v2.0.0).}
    \label{fig:channel_configuration}
\end{figure}

Typically, at least minimal preprocessing of the raw EEG data is applied in both scenarios, relying on handcrafted feature extraction and based on end-to-end approaches. 
We applied the preprocessing steps described below to both scenarios to normalize the input distribution and thus stabilize the deep network learning process, a common practice in deep learning applications, and to stabilize feature extraction.
However, the latter requires additional steps that are described in Section~\ref{sec:feature_based_decoding}.
Importantly, our general preprocessing did not preselect any EEG features.
%as it would be the aim in the hand-crafted feature approach.
As in our earlier work on EEG pathology decoding with deep ConvNets [\cite{schirrmeister2017patho}],
we included the following preprocessing steps: First, we selected a subset of 21 electrode positions (Figure~\ref{fig:channel_configuration}) following the international 10-20 placement [\cite{jasper1958report}] because these electrode positions occurred in all the individual recordings in the data set. 
Then, we discarded the first 60 s of every recording because we observed a large number of recording artifacts in this period, which could have been caused by rearrangement of the electrode cap or by finding a comfortable seating position. Moreover, we used a maximum of 20 min of every recording to avoid considerable feature generation and resampling times for exceptionally long recordings.
As in our previous work \cite{schirrmeister2017patho} and in the work by \cite{van2019detecting}, EEG recordings were downsampled to 100 Hz and clipped at $ \pm 800\mu V $ to reject unphysiologically extreme values and to ensure comparability to these previous studies.
Although \cite{roy2019chrononet} performed their experiments at 250 Hz, we chose to use 100 Hz for better comparability with the other approaches, and to avoid motor artifacts.
However, this could place us at a disadvantage in the direct comparison with \cite{roy2019chrononet}.
Our preprocessing partially uses code from Python libraries MNE\footnote{available for download at \url{https://github.com/mne-tools/mne-python}} and resampy\footnote{available for download at \url{https://github.com/bmcfee/resampy}}.

\subsection{End-to-end decoding with deep neural networks}
\subsubsection{Neural network architectures}
\begin{figure}[ht!]
    \centering
    \includegraphics[width=\textwidth]{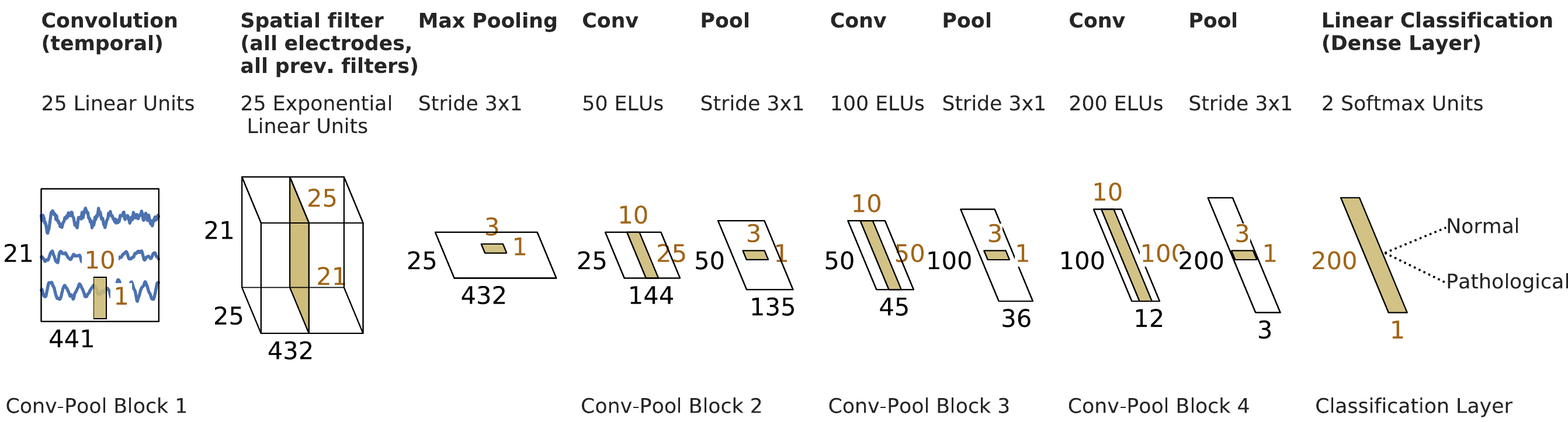}
    \caption{Four-layered BD-Deep4 as introduced by \cite{schirrmeister2017hbm}. Initial separated convolution is followed by several convolution and max-pooling blocks.}
    \label{fig:deep}
\end{figure}

We used different neural network architectures including ConvNets and TCNs to decode the pathology from the EEG recordings. 
First, we used a four-layered ConvNet architecture called BD-Deep4 as previously introduced by \cite{schirrmeister2017hbm}.
The BD-Deep4 architecture [Figure~\ref{fig:deep}] has an initial separated convolution (first temporal, then spatial).
Subsequently, it has several blocks consisting of convolution and max-pooling and uses exponential linear units as activation functions.
It is a rather general architecture that has proven to generalize well to several EEG decoding tasks such as motor (imagery) decoding [\cite{schirrmeister2017hbm}], velocity and speed decoding [\cite{hammer2013role}], and pathology decoding [\cite{schirrmeister2017patho}, \cite{van2019detecting}].
We applied BD-Deep4 without any further adjustments to its architecture.

\begin{figure}[ht!]
    \centering
    \includegraphics[width=\textwidth]{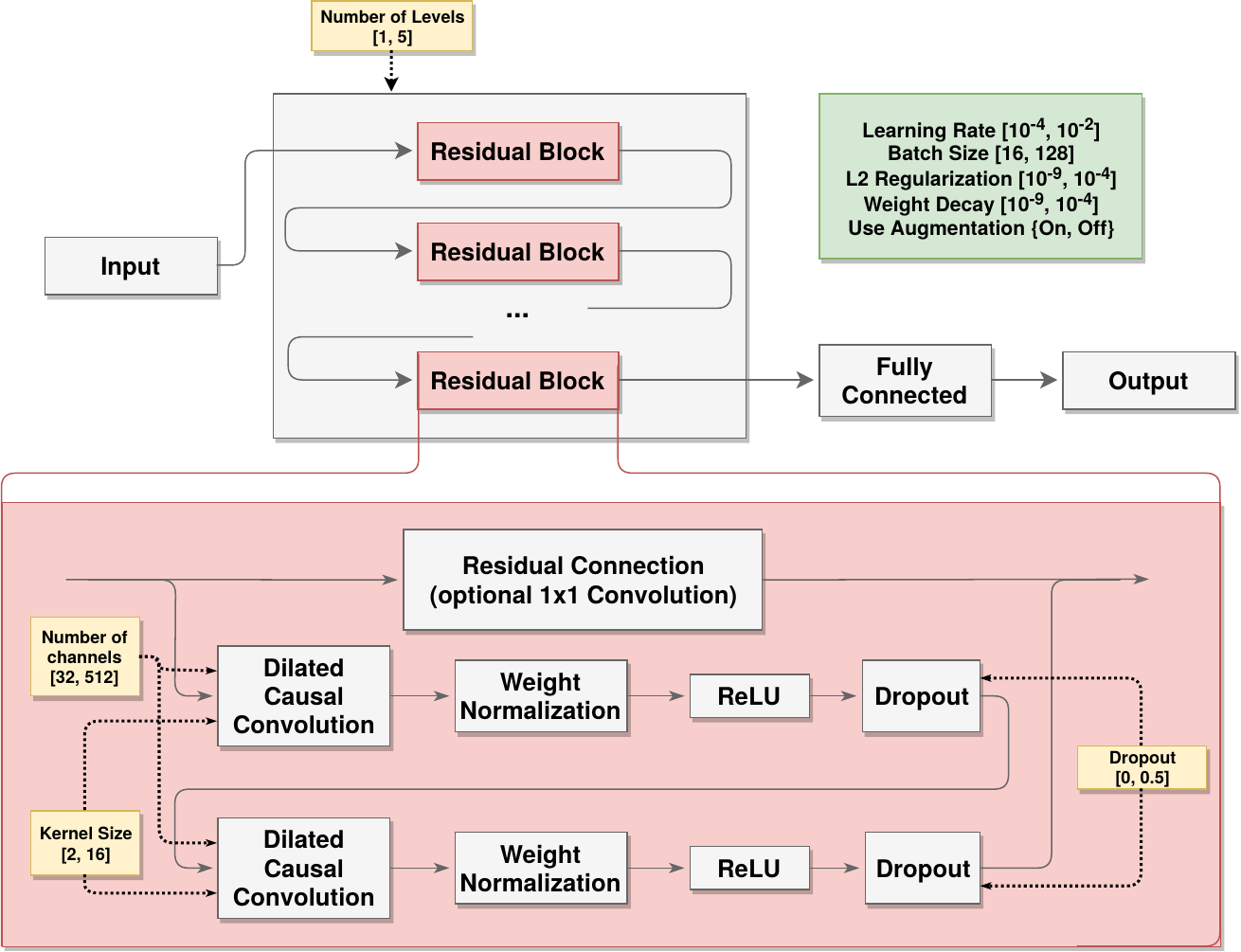}
    \caption{General architecture of TCN as introduced by \cite{bai2018empirical} and search base explored in Master thesis by \cite{chrab2018architecture} to find BD-TCN.}
    \label{fig:tcn}
\end{figure}

Next, we used a TCN architecture [Figure~\ref{fig:tcn}] that was evaluated in a Master thesis by \cite{chrab2018architecture}.
The TCN was originally proposed by \cite{bai2018empirical} as an alternative to recurrent neural networks (RNN) [\cite{rumelhart1988learning}]. 
It is the most complex and deepest architecture under investigation in the present study.
The optimization by \cite{chrab2018architecture} resulted in five levels of blocks consisting of temporal convolutions with 55 channels each as well as max-pooling.
We call this optimized architecture Braindecode TCN (BD-TCN).
For more information on the optimized hyperparameters, see Table~\ref{tab:network_hyperparameters}.

\begin{figure}[ht!]
    \centering
    \includegraphics[width=\textwidth]{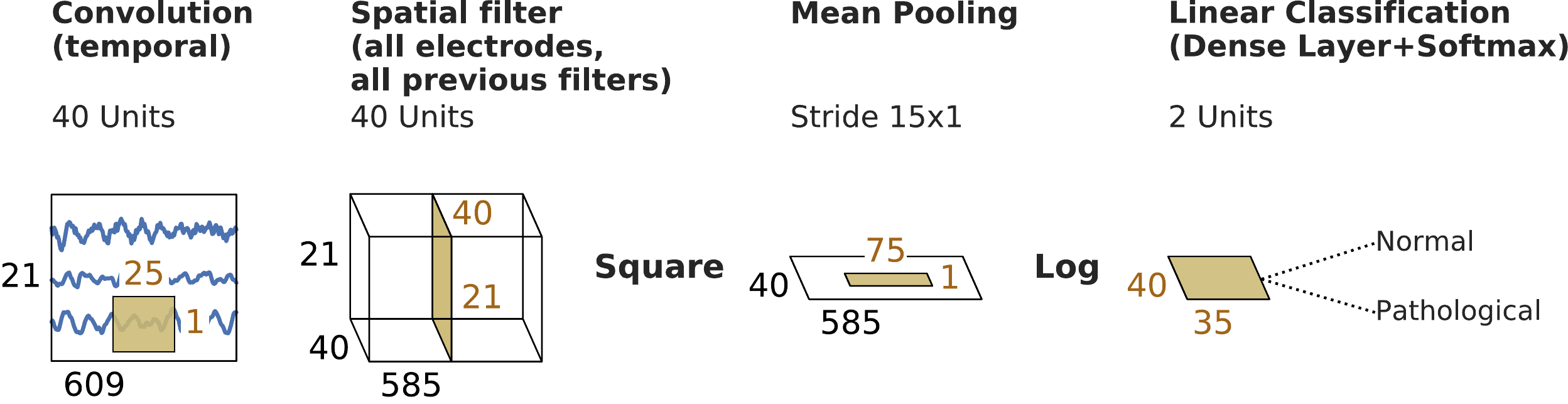}
    \caption{BD-Shallow architecture as originally introduced by \cite{schirrmeister2017hbm} inspired by FBCSP algorithm [\cite{ang2008filter}].}
    \label{fig:shallow}
\end{figure}

Furthermore, we used another ConvNet architecture introduced by \cite{schirrmeister2017hbm} called Braindeocde Shallow ConvNet (BD-Shallow).
The network [Figure~\ref{fig:shallow}], as in the BD-Deep4 network, has an initial separated convolution; however, it is the only convolution in the entire architecture.
The well-known FBCSP algorithm [\cite{ang2008filter}] inspired the BD-Shallow architecture, in particular the squaring and logarithmic nonlinearities. 
It was designed to specifically extract the logarithm of the band power of EEG signals.
We applied BD-Shallow, as BD-Deep, without any further adjustments to its architecture.

Moreover, we used a reimplementation of another ConvNet architecture called EEGNet that was originally introduced by \cite{lawhern2018eegnet}.
We call this reimplementation Braindecode EEGNet (BD-EEGNet).
Again, the architecture has a separated initial convolution. 
Furthermore, the architecture is remarkable owing to its small number of parameters (see Table~\ref{tab:model_parameters}). 

\subsubsection{Training of neural networks}
We trained the networks in a cropped manner with equally-sized, maximally overlapping crops as described by \cite{schirrmeister2017hbm}.
The receptive field of the networks automatically determines the size of the crops.
All networks are exposed to approximately 600 signal samples at a time, except the TCN which has a receptive field of approximately 900 samples [Table~\ref{tab:model_parameters}].
Unlike the original paper by \cite{schirrmeister2017patho}, we used optimizer AdamW [\cite{loshchilov2017fixing}] over Adam [\cite{kingma2014adam}] to minimize the categorical cross-entropy loss function. 
AdamW decouples weight decay updates and the optimization of the loss function, which allows for better generalization [\cite{loshchilov2017fixing}].
We used cosine annealing [\cite{loshchilov2016sgdr}] to schedule the learning rates for the gradient and weight decay updates.
We did not perform learning rate restarts.

\subsection{Feature-based decoding}
\label{sec:feature_based_decoding}

\subsubsection{Additional preprocessing prior to feature extraction}
After general preprocessing common to both pipelines (see Section~\ref{sec:preprocessing}), we applied several additional steps in the feature-based pipeline.
In the special case of connectivity feature extraction, we first filtered entire signals to a selected frequency range in the time domain to avoid the creation of filtering artifacts at the start and end points of the signal segments.
We split every recording into equally sized, nonoverlapping signal segments called crops of 600 samples, i.e., given the sampling frequency of 100 Hz, this corresponded to 6 s, to be maximally comparable to the end-to-end pipeline, where the receptive field of the architectures determines the crop size [\ref{tab:model_parameters}].
We discarded crops with values of $ \pm 800\mu V $ to stabilize feature generation. 
This resulted in the exclusion of one recording (subject 00008184, session s001\_t001), as one channel was not properly recorded, meaning each measurement at every time point exceeded the outlier value. 

\subsubsection{Feature extraction}
We computed a large set of features describing time, frequency, and connectivity structure of the EEG signals that have all been used to characterize EEGs [\cite{subasi2007eeg}, \cite{logesparan2012optimal}, \cite{kuhlmann2008correlation}, \cite{kumar2010entropies}, \cite{quiroga1997searching}, \cite{hjorth1970eeg}, \cite{james2003extracting}, \cite{petrosian1995kolmogorov}, \cite{inouye1991quantification}, \cite{roberts1999temporal}, \cite{balli2009combined}, \cite{peng1995quantification}, \cite{watter2014epileptic}, \cite{minasyan2010patient}, \cite{van2005detecting}, \cite{esteller2001line}, \cite{katz1988fractals}, \cite{lachaux1999measuring}]. 
We generated features on every crop based on Discrete Fourier Transform (FT), Continuous and Discrete Wavelet Transform (CWT and DWT), and between-electrode connectivity features based on the Hilbert Transform [Table~\ref{tab:features}].
Furthermore, we parsed age and gender of the patients from the European Data Format [\cite{kemp1992simple}] recording file headers as optional additional features. 
The feature implementations partially use code from the Python libraries PyEEG\footnote{\label{note:pyeeg}available for download at \url{https://github.com/forrestbao/pyeeg}} and  PyWavelets\footnote{\label{note:pywavelets}available for download at \url{https://github.com/PyWavelets/pywt}}.
The implementations can be found in our feature decoding toolbox \href{https://github.com/TNTLFreiburg/brainfeatures}{Brainfeatures}\footnote{\label{ntoe:brainfeatures}available for download at \url{https://github.com/TNTLFreiburg/brainfeatures}}. 

\begin{table}[ht!]
    \centering
    \begin{tabular}{ccc} 
        \multicolumn{1}{c|}{\textbf{CWT/DWT$^+$}} & \multicolumn{1}{c|}{\textbf{FT}} & \multicolumn{1}{c}{\textbf{Patient information}} \\
        \multicolumn{1}{c|}{Bounded variation} & \multicolumn{1}{c|}{Maximum} & \multicolumn{1}{c}{Age} \\
        \multicolumn{1}{c|}{Maximum} & \multicolumn{1}{c|}{Mean} & \multicolumn{1}{c}{Gender} \\ 
        \multicolumn{1}{c|}{Mean} & \multicolumn{1}{c|}{Minimum} & \multicolumn{1}{c}{} \\ \cline{3-3}
        \multicolumn{1}{c|}{Minimum} & \multicolumn{1}{c|}{Peak frequency} & \multicolumn{1}{c}{\textbf{Riemannian}} \\       
        \multicolumn{1}{c|}{Power} & \multicolumn{1}{c|}{Power} & \multicolumn{1}{c}{Covariance matrix} \\ 
        \multicolumn{1}{c|}{Power ratio} & \multicolumn{1}{c|}{Power ratio} & \multicolumn{1}{c}{} \\ \cline{3-3}
        \multicolumn{1}{c|}{Spectral entropy} & \multicolumn{1}{c|}{Spectral entropy} & \multicolumn{1}{c}{\textbf{Connectivity}} \\       
        \multicolumn{1}{c|}{Variance} & \multicolumn{1}{c|}{Value range} & \multicolumn{1}{c}{Phase Locking Value} \\       
        \multicolumn{1}{c|}{} & \multicolumn{1}{c|}{Variance} & \multicolumn{1}{c}{} \\ \hline
        \textbf{Time} & Hjorth mobility & Minimum \\
        Detrended Fluctuation Analysis* & Hurst exponent & Nonlinear energy\\
        Energy & Kurtosis & Petrosian fractal dimension*\\
        Fisher information* & Line length & Skewness\\
        Fractal dimension & Lyauponov exponent* & SVD entropy*\\
        Higuchi fractal dimension & Maximum & Zero crossings\\
        Hjorth activity & Mean & Zero crossings of derivative\\
        Hjorth complexity & Median &
    \end{tabular}
    \caption{
    All implemented features sorted by feature domain. Feature domains are CWT/DWT, FT, Patient information, RG, Connectivity, and Time. Features marked with * were computed using \href{https://github.com/forrestbao/pyeeg}{PyEEG}. Features marked with $^+$ were computed using \href{https://github.com/PyWavelets/pywt}{PyWavelets}.}
    \label{tab:features}
\end{table}

For the CWT, DWT, and FT feature computation, we weighted time domain crops with a Blackman-Harris window function to enhance the spectral estimation and reduce the effect of leakage. 
In preliminary experiments [\cite{gemein2017automated}], we tested different window functions. 
The Blackman-Harris window yielded best results, although the choice of the window function had only minor effects on the decoding accuracy. 
We extracted frequency features from bands 0--2 Hz, 2--4 Hz, 4--8 Hz, 8--13 Hz, 13--18 Hz, 18--24 Hz, 24--30 Hz, and 30--50 Hz using FT and a band overlap of 50\%. 
We chose the bands to match the frequency bands commonly used in the literature.
Furthermore, we observed superior results in preliminary experiments when using a band overlap [\cite{gemein2017automated}].
We chose wavelet scales for CWT and levels for DWT to match these bands as closely as possible. 

For connectivity feature computation, we transformed the frequency-filtered time-crops using the Hilbert Transform to extract the signal phase.

The dimension of the feature vectors, including all feature values of domains CWT, DWT, FT, Connectivity, and Time, was $ F=8631 $.

\paragraph{Time-resolved features}
Feature generation resulted in a feature matrix $ {M}_{i} \in \mathbb{R}^{C_{i} \times F} $ for every recording, where $ I $ is the total number of recordings, $ C_{i} $ is the number of analyzed 6-s crops $ i \in I $, and $ F $ is the dimension of the feature vector. 
For time-resolved (non-aggregated) decoding, we considered every feature vector of every crop as an independent example. 
This drastically increased the number of training examples, which could be beneficial in the training phase. 
However, it also resulted in higher memory consumption and higher learning times. 

\paragraph{Aggregated features}
For aggregated decoding, we computed the aggregate of all time-crop feature matrices $ M_{i} \in \mathbb{R}^{C_{i} \times F} $. 
Therefore, we used the median as the aggregation function, such that we obtained a single feature vector of length $ F $ for each recording. 
In previous experiments [\cite{gemein2017automated}], the median proved to be the best aggregation function in terms of decoding accuracy, although, again, the choice only had a minor effect.
Aggregation drastically reduces the feature matrix size, which allows for faster learning and prediction. 
However, it has the disadvantage of discarding all time-resolved information as it collapses features of all crops of a recording into a single feature vector. 
The shape of the final aggregated feature matrix was $ M_{aggregate} \in \mathbb{R}^{I \times F} $.

\paragraph{Dimensionality reduction}
We reduced the feature dimension $F$ in preliminary experiments using principal component analysis (PCA) [\cite{wold1987principal}]; however, independent of the choice of principal components or ratio of variance, the application of PCA led to a decrease in decoding accuracy.
 
\paragraph{Covariance matrices} 
For Riemannian-geometry-based decoding (see below) we computed a covariance feature matrix $ \Sigma_{i} \in \mathbb{R}^{C_i \times E*E} $ for every crop, where $E$ is the number of electrodes. 
Therefore, we used the Python package pyRiemann \footnote{\label{note:pyriemann}available for download at \url{https://github.com/alexandrebarachant/pyRiemann}}.
We independently tested the Euclidean and the geometric means to aggregate covariance matrices of the crops, such that we obtained a feature vector of length $ E * (E + 1)/2 $ for each recording and aggregation type. 
The shape of the final covariance feature matrix was $ M_{riemann} \in \mathbb{R}^{I \times E * (E + 1)/2}$.

\subsubsection{Feature-based classifiers}
After feature generation, we used the feature matrix as an input to several feature-based ML models. 
We used an SVM with radial basis function (RBF) kernel, as commonly used in the literature [\cite{lehmann2007application}, \cite{cai2016pervasive}].
Furthermore, we used an RF classifier which is, by design, robust towards overfitting and hence a reliable baseline model.
Furthermore, we also applied the automated ML toolkit ASC\footnote{\label{note:autosklearn}available for download at \url{https://github.com/automl/auto-sklearn}} as it has the potential to yield superior results owing to the automatic ensemble selection and hyperparameter optimization. 
For more information on this toolkit see \cite{feurer2015efficient}. 
Finally, we evaluated the Riemannian-geometry-based decoding as implemented in the Python package pyRiemann\footnote{see footnote \ref{note:pyriemann}} using an SVM with an RBF kernel, as it has recently achieved state-of-the art results in several BCI decoding tasks [\cite{lotte2018review}].
All models under investigation relied on implementations in \href{https://scikit-learn.org/stable/}{scikit-learn}\footnote{\label{note:sklearn}available for download at \url{https://scikit-learn.org/stable/}} [\cite{pedregosa2011scikit}]. 

\subsection{Evaluation of performance}
We performed 5-fold cross-validation (CV) on the development set, such that each recording in the development set was predicted exactly once. 
We did not shuffle data during splitting; rather, we used chronologically ordered splits.
During CV, we optimized the hyperparameters of our feature-based models [see Table~\ref{tab:feature_hyperparameters}]. 

For final evaluation, we evaluated our models on the held back final evaluation set. 
We trained models on the full development set and predicted the examples in the final evaluation set. 
We repeated final evaluation five times to manage the statistical variances caused by initialization of certain models. 

We report the accuracy score as $ ACC = \frac{\text{\# of correct predictions}}{\text{\# of examples}} $ for the development and final evaluation sets as averages over the CV folds and final evaluation repetitions, respectively.

We used a statistical sign test [\cite{dixon1946statistical}] to validate the predicted labels in the final evaluation for superiority of model performance ($ H1 $). 
To provide a conservative estimate, occurring ties were equally split to both classes.
We rejected the null hypothesis ($ H0 $: There is no difference in performance) at a p-value \textless $0.05$.

\subsection{Analysis}
\paragraph{Handcrafted features}
Because we implemented a large set of features [Table~\ref{tab:features}], we were interested in their individual importance to the decision process. 
RF estimates the importance of a feature by internally computing the ``purity'' of a data split obtained through a feature.
In principal, the purer the data split and the earlier the feature is considered in the trees of the forest, the higher its importance.
We assigned a textual label to all computed features described in Section~\ref{sec:feature_based_decoding} and mapped them to the average feature importance in CV using RF.
We selected subsets of the features based on their textual labels with respect to frequency band and electrode location.
We then created topological plots of the average feature importance in certain frequency ranges.
Furthermore, we computed the Spearman correlation of features over the development set and visualized the correlation map, as pronounced correlations could be a limiting factor to the interpretation of feature-based analyses.

\paragraph{Learned features}
We performed input-signal perturbation to determine the informative frequency ranges and electrode locations for identifying pathologies in the EEGs [\cite{schirrmeister2017hbm}]. 
We computed the network predictions of the original and randomly perturbed input signals and correlated the amplitude change with the change of predictions.
Given the labels of the examples, we could then determine whether an increase (or decrease) of signal amplitude in a given frequency range through perturbation contributed to more pathological (or non-pathological) predictions.
Again, we made topological plots to indicate the most correlated frequency range and electrode location with the pathological class.

\subsection{Ensembling}
In ensembling, a number of model predictions are combined to provide an improvement over single-model performances.
Under the assumption that models make uncorrelated errors, a combination can result in the overruling of incorrect single-model decisions.
We computed the Spearman correlation of the CV predictions of all pairs of models and visualized the resulting correlation map. 
Furthermore, we computed and visualized the ratio of non-overlapping label errors of pairs of models to investigate the possibility of ensembling.
Note that ASC was excluded from this investigation because it does not provide access to internally performed CV predictions.
For ensembling, predictions are weighted.
A weighting of ``1'' of individual model predictions is a special case and is equivalent to majority voting.
We first built an ensemble of three models based on the highest ratio of non-overlapping label errors.
We then computed an ensemble label based on the majority vote of labels computed from the individual model predictions.
In addition, as an automated alternative, we investigated an ensemble selection technique based on \href{https://github.com/automl/auto-sklearn}{auto-sklearn}\footnote{see footnote \ref{note:autosklearn}} introduced by \cite{caruana2004ensemble} that automatically selects models for ensembling and computes the optimal weights based on the validation set.
We evaluated the performance of both ensembles based on CV and final evaluation predictions of the individual models.

\section{Results}
\label{sec:results}
\subsection*{Data descriptive statistics}
\label{sec:result0}
\begin{figure}[ht!]
  \centering
  \begin{minipage}{0.49\textwidth}
    \includegraphics[width=\textwidth]{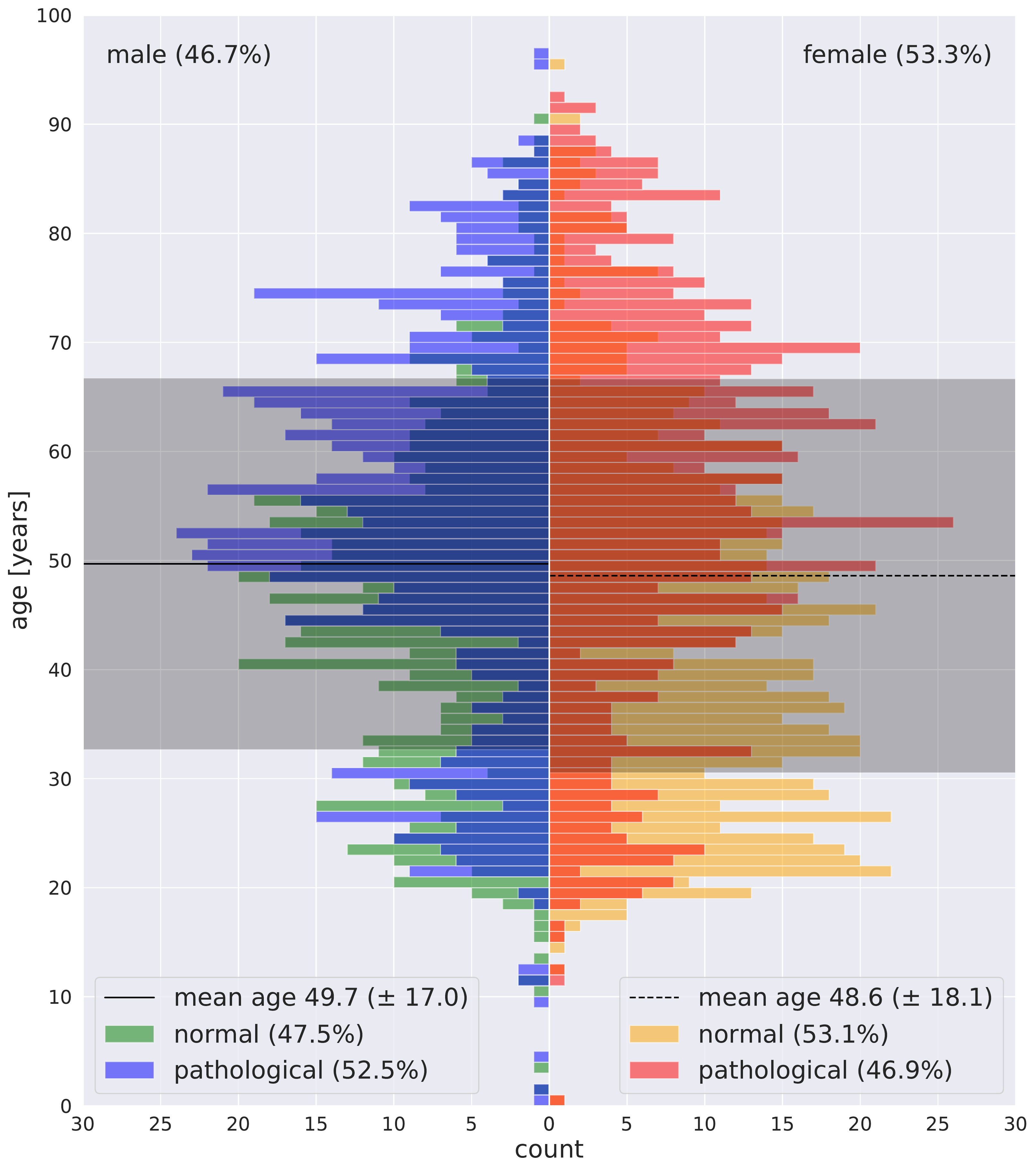}
  \end{minipage}
  \hfill
  \begin{minipage}{0.49\textwidth}
    \includegraphics[width=\textwidth]{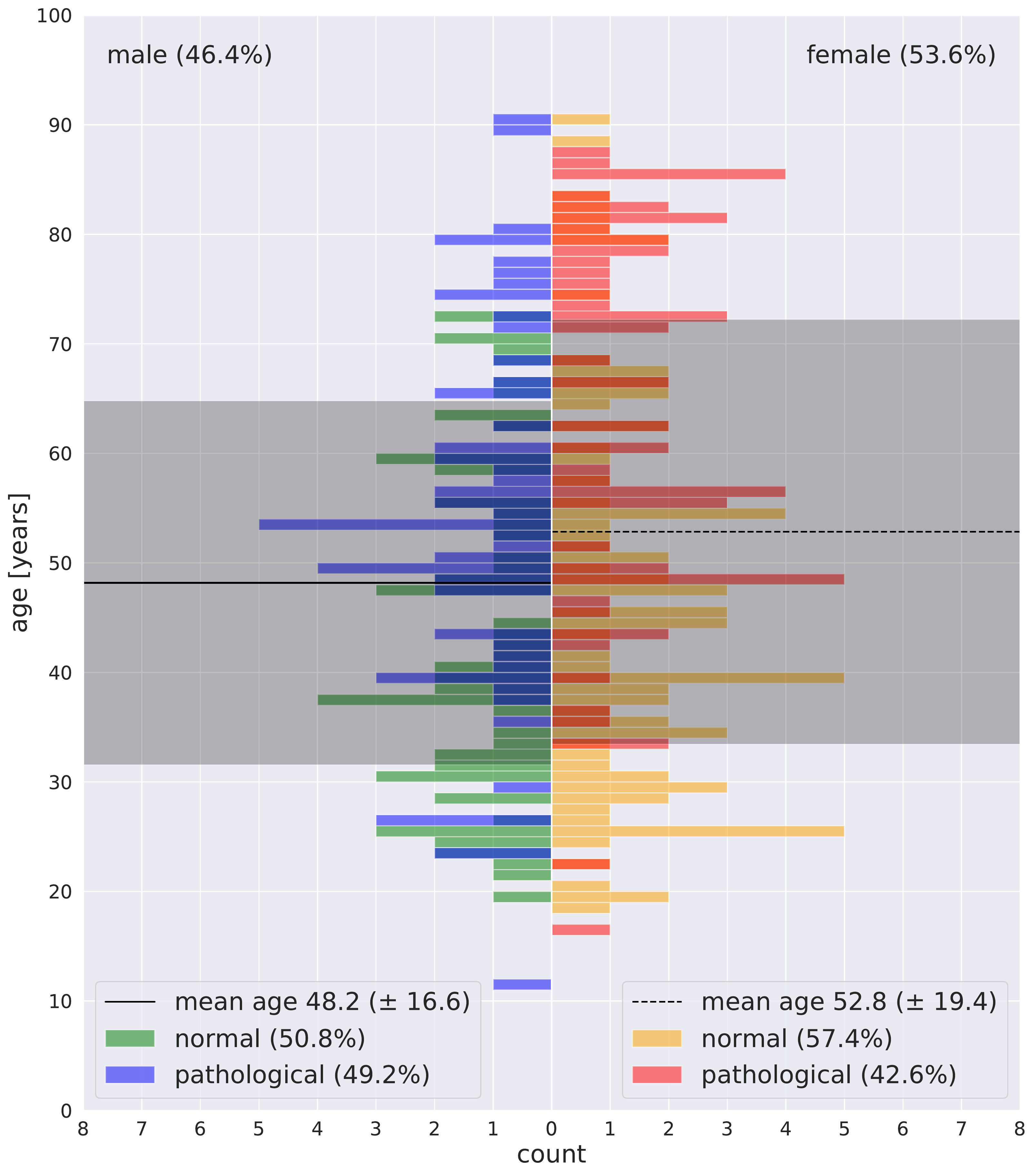}
  \end{minipage}
  \caption{Development (left) and final evaluation (right) subsets of TUH Abnormal EEG Corpus (v2.0.0). Histogram is constructed as an age pyramid subdivided into male and female patients. Different color coding indicates pathological and non-pathological EEG recordings.}
  \label{fig:tuh_histogram}
\end{figure}

We present histograms of the age distribution within the development and final evaluation sets of the TUH Abnormal EEG Corpus (v2.0.0) in Figure~\ref{fig:tuh_histogram}.
The age distribution, especially of the female patients, differed between the development and final evaluation sets. 
Moreover, the ratio of pathological and non-pathological examples also differed by gender and subset. 
The proportion of male and female patients, conversely, was closely matched.
It can be observed that recordings labeled as pathological appear more frequent with higher age, which matches the intuition.

The observations are important in two respects. 
First, correlation of pathology with age could lead to a situation where the trained models use patient age as a proxy for pathology. 
To investigate the role of patient age in the decoding of EEG pathology further, we included age as a feature in a separate analysis (Section~\ref{sec:result6}). 
Conversely, systematic differences between the development and final evaluation sets can reduce generalization and thus present a challenge. 
However, both correlation of pathology with age and shifts between the development and final evaluation data could occur in practical application scenarios. We hence considered these properties of the TUH data set as ecologically valid and methodologically interesting aspects.
However, they must be considered when interpreting the results achieved on this data set.   

\subsection{Aggregated feature-based decoding}
\label{sec:result1}
\begin{figure}[ht!]
    \centering
    \includegraphics[width=\textwidth]{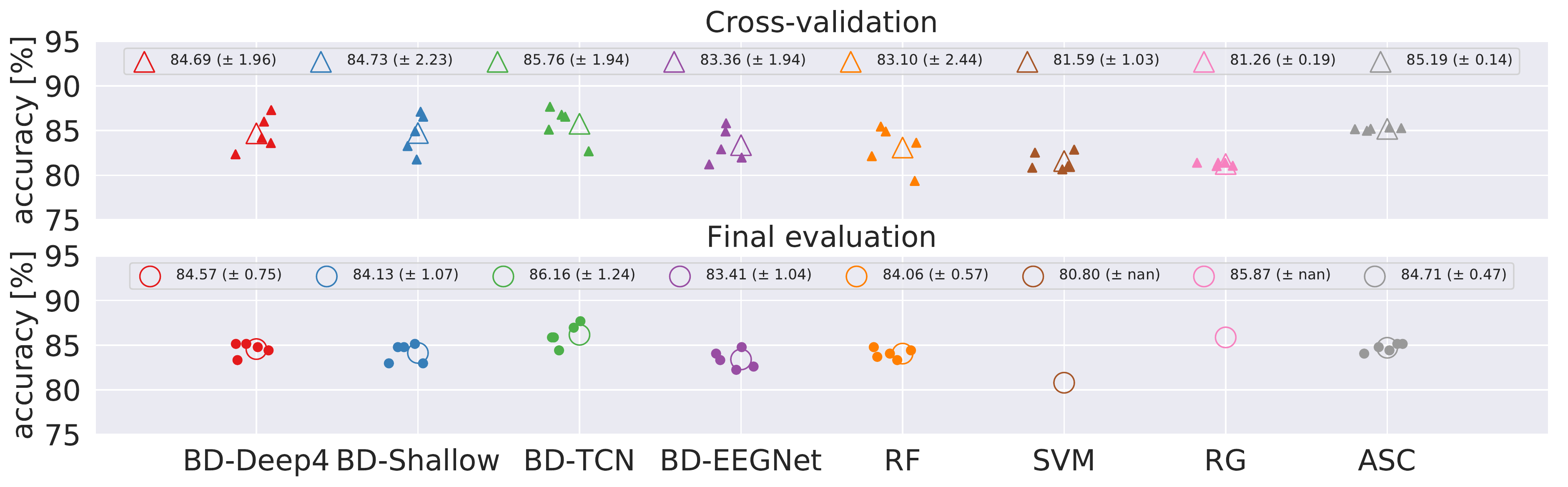}
    \caption{Decoding accuracies of all models during CV and in final evaluation. TCN implemented in Braindecode (BD-TCN) indicated best performance.
    Decoding based on RG achieved accuracy similar to BD-TCN.
    BD-Deep4 and BD-Shallow ConvNets, RF, and ASC were on same level, whereas BD-EEGNet achieved marginally lower decoding accuracy. 
    SVM indicated the worst performance.}
    \label{fig:result_overview}
\end{figure}

We present the aggregated feature-based decoding results in Figure~\ref{fig:result_overview}, right half.
The Riemannian-geometry-based pathological versus non-pathological EEG decoding achieved nearly 86\% accuracy. 
Interestingly, the decoding accuracy increased from 81\% in CV to 86\% in the final evaluation, which could indicate underfitting of the training data in CV.
We obtained greater than 84\% accuracy using both a traditional and automated feature-based approach. 

\begin{figure}[ht!]
    \centering
    \includegraphics[width=\textwidth]{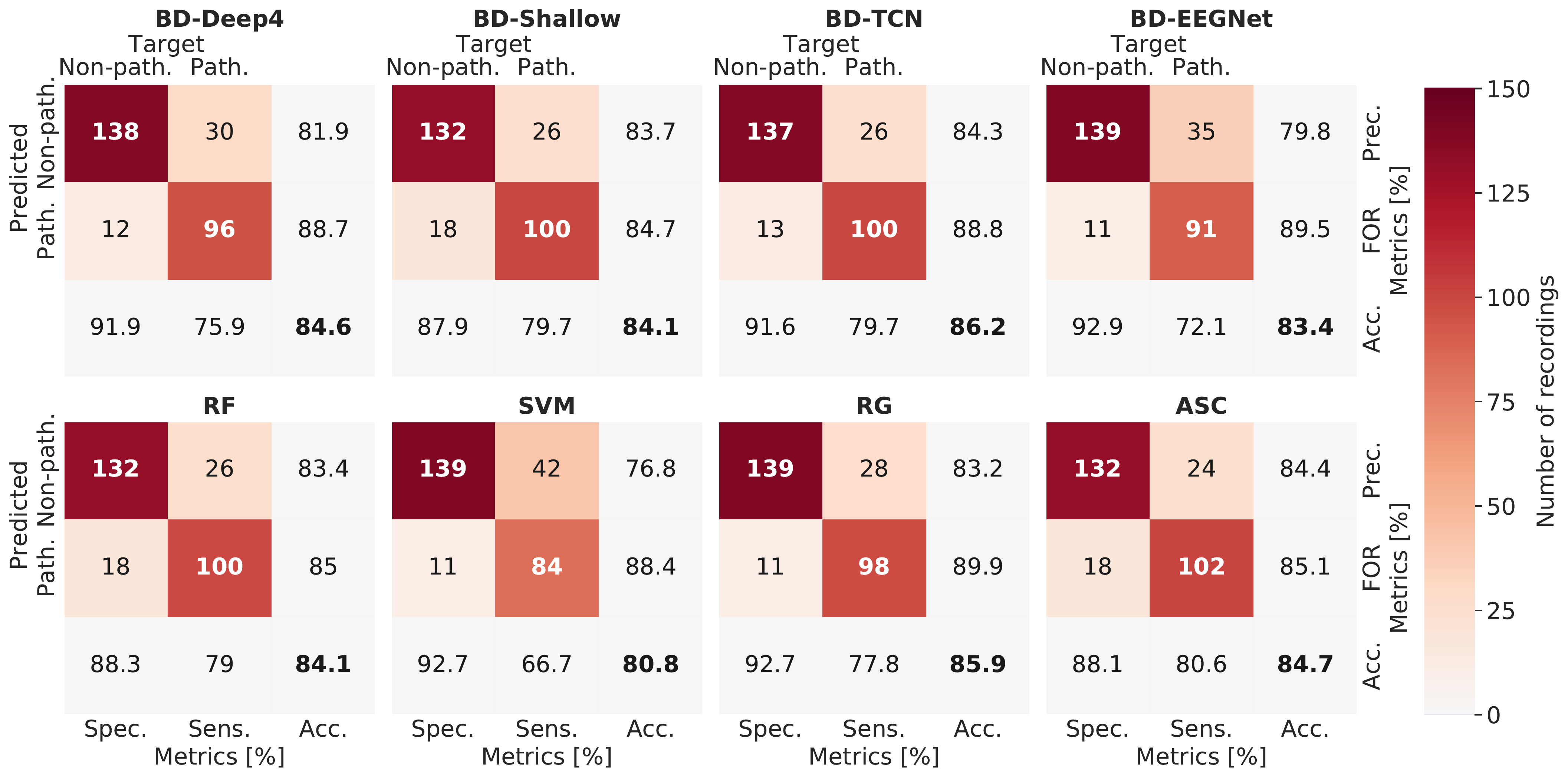}
    \caption{Confusion matrices of all models averaged over independent final evaluation runs are presented in the upper left $2 \times 2$ sub-matrix. Specificity (Spec.), sensitivity (Sens.), precision (Prec.), false omission rate (FOR), and accuracy (Acc.) are indicated. All models determined more false negatives (pathological examples predicted as non-pathological).}
    \label{fig:confusion_matrix}
\end{figure}

Furthermore, we observed that all feature-based models had a higher ratio of false negatives than false positives; that is, they rather classified pathological examples as non-pathological than the opposite (Figure~\ref{fig:confusion_matrix}, bottom row).
This is in consensus with results presented by \cite{abnormalLopez}, \cite{schirrmeister2017patho}, and \cite{van2019detecting}.

To the best of our knowledge, there is only one other previously published feature-based result for pathology decoding based on the TUH Abnormal EEG corpus [\cite{abnormalLopez}].
It was achieved using cepstral coefficients and a CNN+MLP architecture for classification resulting in an accuracy of 78.8\%.
We thus increased this feature-based baseline by greater than 5\% using RF and ASC, and more than 7\% using RG. 

Riemannian-geometry-based classification outperformed all other feature-based models and achieved 85.87\% accuracy.
We observed that treating covariance matrices with appropriate metrics of their native space (geometric instead of Euclidean mean) yielded superior performance [\ref{tab:riemann_comparison}], which was to be expected [\cite{barachant2013classification}].

This performance of the Riemannian-geometry-based decoding was remarkable considering that the covariance matrix of 21 electrodes had only 231 non-redundant entries.
The covariance matrices, as well as our aggregated high-dimensional feature vectors, did not contain detailed time course information, and, in both cases, we averaged over the number of crops from which features were extracted. 
However, the results demonstrate that there is sufficient information contained in covariance matrices even to outperform all other tested models using handcrafted features. 

The application of ASC effectively reduces the required amount of time and expert knowledge required to build and optimize a well-working model and has the ultimate goal of making ML applicable by non-experts.
We can confirm that, given our set of features, ASC achieved competitive results in classifying EEG pathology, without requiring user interaction.
The ensemble that was automatically chosen by ASC consisted of AdaBoost [\cite{schapire2013boosting}] with 78\% (66\%, 8\% and 4\%) of ensemble strength, gradient boosting with 18\%, and linear discriminant analysis with 4\%.

\subsection{End-to-end decoding performance}
\label{sec:result2}
We present the end-to-end decoding results with deep neural networks in Figure~\ref{fig:result_overview}, left half.
Our overall best decoding result was 86.16\% accuracy obtained by the BD-TCN.
This accuracy was extremely close to the result of 86.57\% accuracy previously reported using ChronoNet [\cite{roy2019chrononet}]. 
The BD-TCN was followed by BD-Deep4 and BD-Shallow with 84.57\% and 84.13\% accuracy, respectively.
BD-EEGNet achieved a decoding accuracy of 83.41\%.
Overall, the networks did not demonstrate as much performance difference in the CV and final evaluation as the feature-based approaches.
For the networks, the differences were in the range of $ -0.6\% $ to $+0.4\%$ and in the range of $-1.21\%$ to $+4,61\%$ for the feature-based approaches.
Again, as in the feature-based approach, we observed more false negatives (Figure~\ref{fig:confusion_matrix}, top row).

\begin{figure}[ht!]
    \centering
    \includegraphics[width=\textwidth]{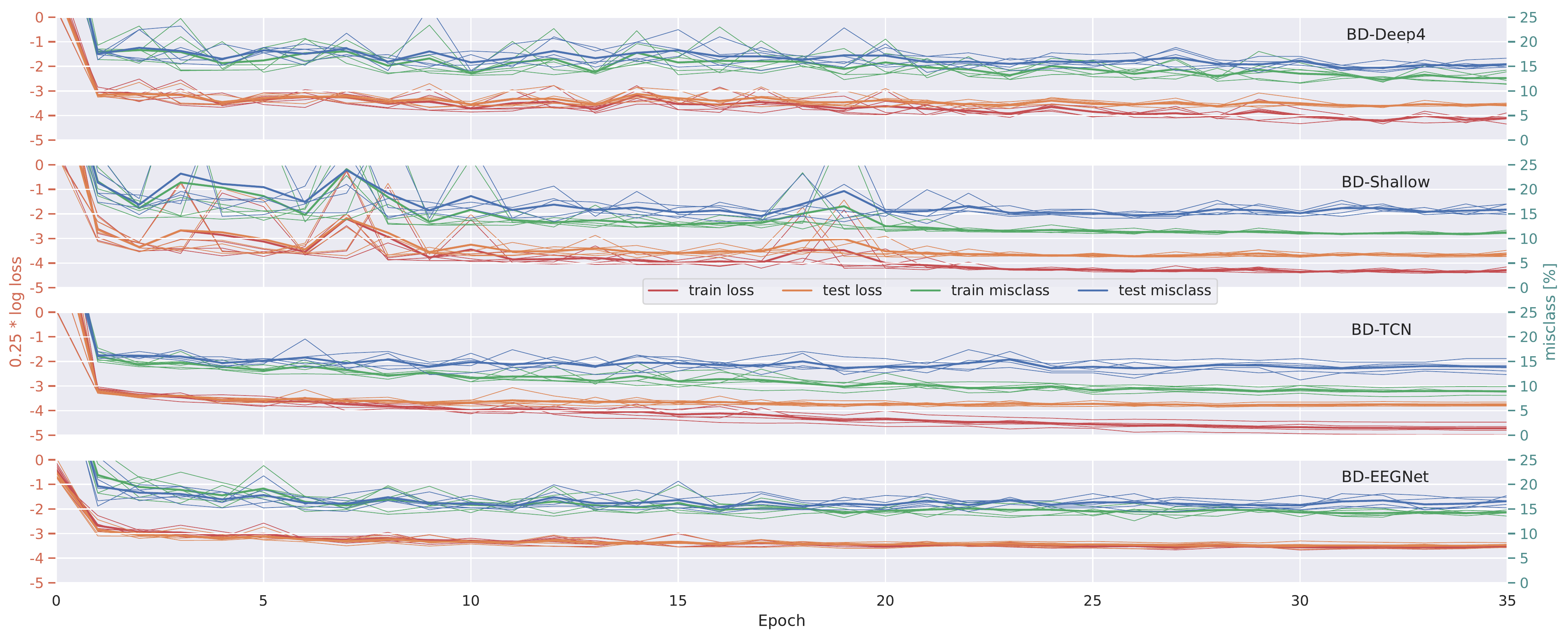}
    \caption{Learning curves of investigated network architectures. Smoothing trend can be observed near Epoch 20, which could be effect of cosine annealing. BD-TCN achieves lowest misclassification rate. Curves of BD-EEGNet are a sign of underfitting.}
    \label{fig:learning_curves}
\end{figure}

We also tracked learning curves of all applied networks (Figure~\ref{fig:learning_curves}).
The loss and misclassification curves of BD-Deep4 and BD-Shallow are irregular at the start; however, they become smoother after Epoch 20. 
BD-EEGNet displayed smooth curves overall, however with the highest loss and misclassification rate. 
The curves of the BD-TCN indicate the greatest difference of training to test loss. 

The BD-TCN was designed as an alternative to the RNN architectures. 
Our results indicate that this model is indeed competitive with ChronoNet, which is a combined ConvNet/RNN architecture, for the given task in terms of decoding accuracy. 
The BD-TCN outperformed all other networks, which could be a consequence of its design and optimization through a neural architecture search [\cite{chrab2018architecture}]. 
All other models were originally developed and optimized for other decoding tasks.
Their performance in the present study underlines their general applicability. 

The presented learning curves demonstrate astounding differences and we hypothesize that they are characteristic for the network architectures under investigation.
The smoothing of the curves near Epoch 20, especially observable for BD-Deep4 and BD-Shallow, is likely the effect of cosine annealing updates of the learning rate. 
In all models, except BD-EEGNet, we can observe a clear difference between the training and test loss.
We assume that BD-EEGNet is unable to better fit the training data owing to its relatively small number of parameters (see Table~\ref{tab:model_parameters}); its learning curves indeed indicate signs of underfitting. 
The opposite, overfitting, cannot be observed. 
This could be due to the regularization techniques, i.e., dropout or weight decay.

\subsection{Feature-based and end-to-end decoding performances in comparison}
\label{sec:result3}
\begin{figure}[ht!]
  \centering
  \includegraphics[width=.8\textwidth]{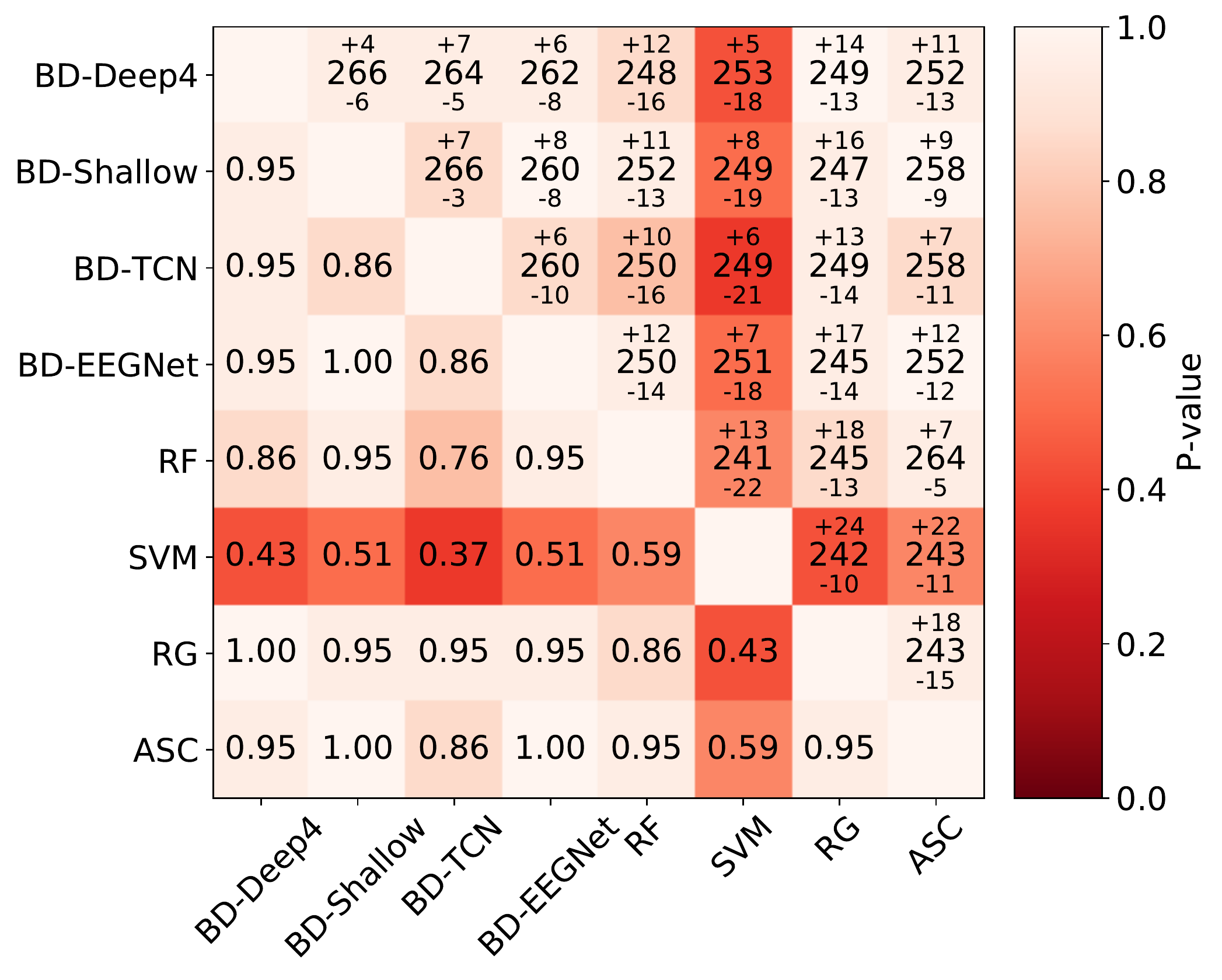}
  \caption{Results of paired sign tests with equally split ties. Lower triangle indicates p-values for all pairs of models. The upper triangle displays the number of ties and number of positive and negative signs. Statistically, there is no evidence to assume that any one model performed better than the other models.}
  \label{fig:sign_test}
\end{figure}

We observed decoding accuracies in the same range using different models and approaches, i.e., ChronoNet (86.57\%), BD-TCN (86.16\%), BD-Deep4 (84.57\%), ASC (84.71\%), RF (84.06\%), and RG (85.87\%). 
Our statistical analysis indicated p-values in the range of 0.37 to 1.
There is thus no statistical evidence that one of these models performed better than the others (Figure~\ref{fig:sign_test}, lower triangle). 
Over all pairs of models, the number of equal classification decisions (ties) ranged from 241--266 of 276 possible (Figure~\ref{fig:sign_test}, upper triangle). 
Several pairs of models actually provided an identical number of errors, i.e., BD-EEGNet, BD-Shallow, and ASC.

Previous publications [Table~\ref{tab:related_works}] have indicated that deep learning typically performs better in decoding pathology from EEGs using the TUH Abnormal EEG corpus.
This is because there is only one published result using handcrafted features.
It appears that this baseline was not particularly strong, which made deep learning approaches appear superior.
However, herein, we determined similar decoding accuracies of feature-based and neural network approaches.
Statistically, there is actually no evidence that one of the models applied for this study performed better than the others .
Hence, we conclude that the literature provided a false picture of reality caused by the omission of comparisons of deep learning with other approaches.

\subsection{Importance of learned and handcrafted features}
\label{sec:result4}
\begin{figure}[ht!]
  \centering
  \includegraphics[width=\textwidth]{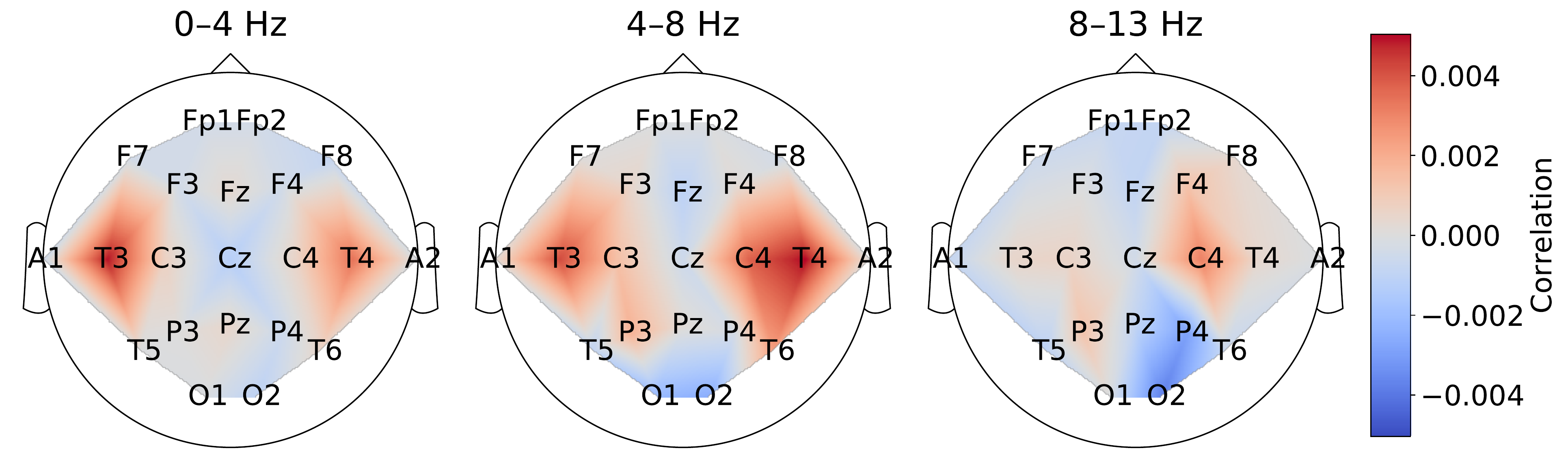}
  \caption{Input-signal perturbation with BD-Deep4 network displays that higher activity at temporal electrode locations (T3, T4) is indicative of pathology, especially in low frequency ranges (0--4 Hz and 4--8 Hz). In alpha frequency range, there is negative correlation with pathological class at occipital electrode locations (O1, O2).}
  \label{fig:perturbation}
\end{figure}

Through our perturbation analysis using the BD-Deep4, we determined correlations with predictions of the pathological class at temporal electrode locations T3 and T4 when increasing the amplitudes (Figure~\ref{fig:perturbation}). 
The effect was especially prominent in the delta and theta frequency range. 
Conversely, a decrease in correlation at occipital electrodes O1 and O2 was the most prominent effect in the alpha frequency range. 

\begin{figure}[ht!]
  \centering
  \includegraphics[width=\textwidth]{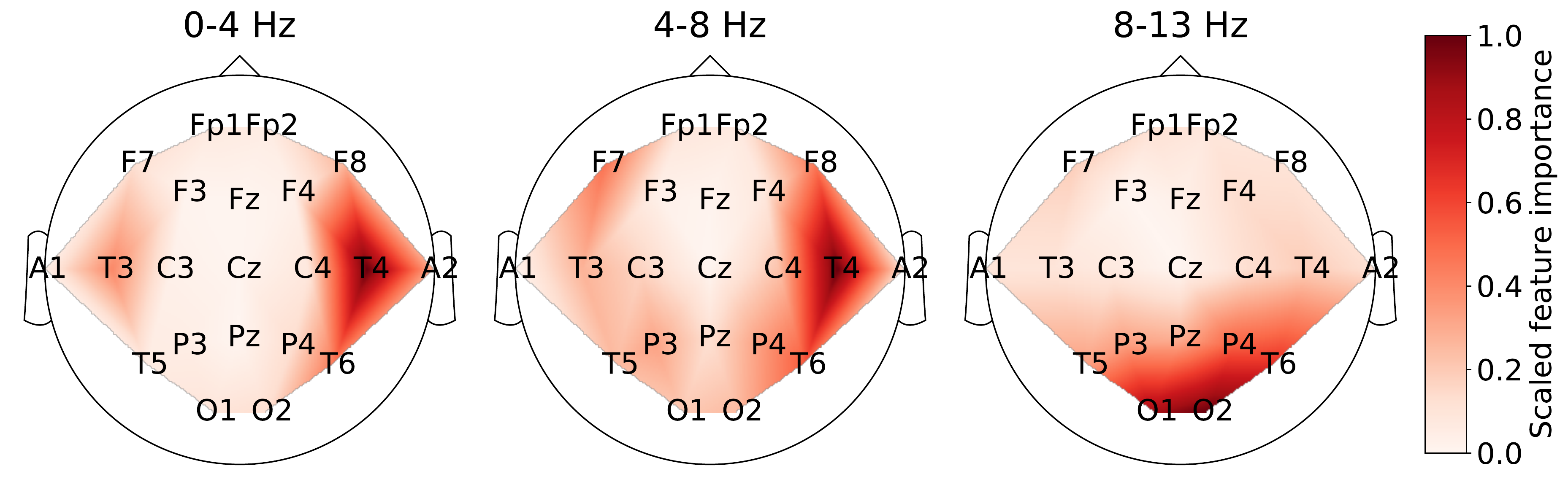}
  \caption{RF feature importance analysis indicates that features extracted at 0--4 Hz and 4--8 Hz at temporal electrode T4 are most informative. In 8--13 Hz frequency band, occipital electrodes (O1, O2) have highest importance values.}
  \label{fig:important_electrodes}
\end{figure}

For a comparison with the perturbation analysis, we present the important handcrafted features extracted in the same frequency ranges using our analysis of feature importance using RF (Figure~\ref{fig:important_electrodes}).
Features extracted at temporal electrode T4 in the delta and theta frequency range are most informative, which is in consensus with the perturbation result.
Features extracted at electrode T3, however, are not considered as informative. 
In the alpha frequency range, the majority of the informative features were extracted at occipital electrodes O1 and O2. 
This, again, is in consensus with the perturbation result.

\begin{figure}[ht!]
  \centering
  \includegraphics[width=\textwidth]{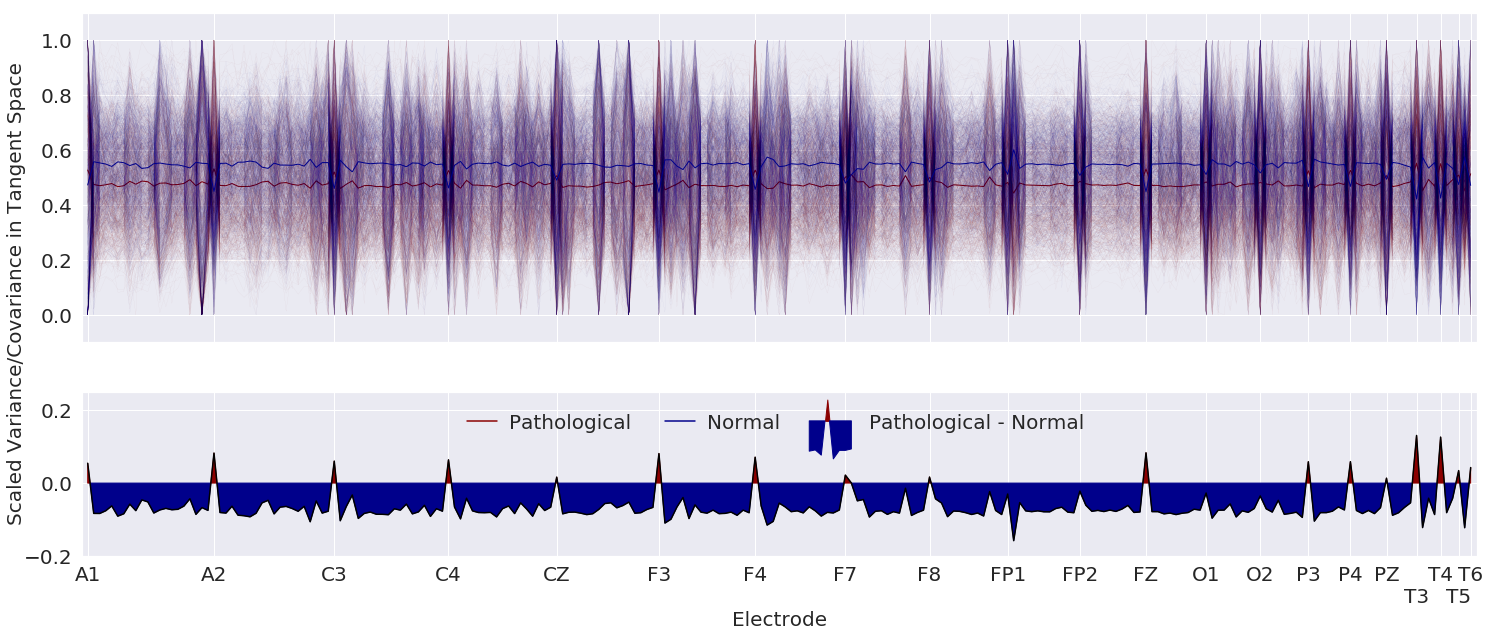}
  \caption{Values of covariance matrices mapped to tangent space. Values most indicative of pathological activity are variance extracted at electrodes T3 and T4. Only variance of O1, O2, Fp1, and Fp2 is indicative of normal activity.}
  \label{fig:covariances_in_ts}
\end{figure}

For a comparison with the perturbation and feature importance analysis, we present a visualization of the values in the covariance matrices that were mapped to tangent space as used in the Riemannian-gemoetry-based decoding pipeline (Figure~\ref{fig:covariances_in_ts}).
It can be observed that the variance extracted at temporal electrodes T3 and T4 is most indicative of a pathology. 
This is in consensus with the perturbation analysis.
Furthermore, as in the perturbation analysis, variance at electrodes O1 and O2 (and Fp1 and Fp2) are indicative of normal brain activity.
Both observations are underlined by the feature importance analysis.

\begin{figure}[ht!]
  \centering
  \includegraphics[width=\textwidth]{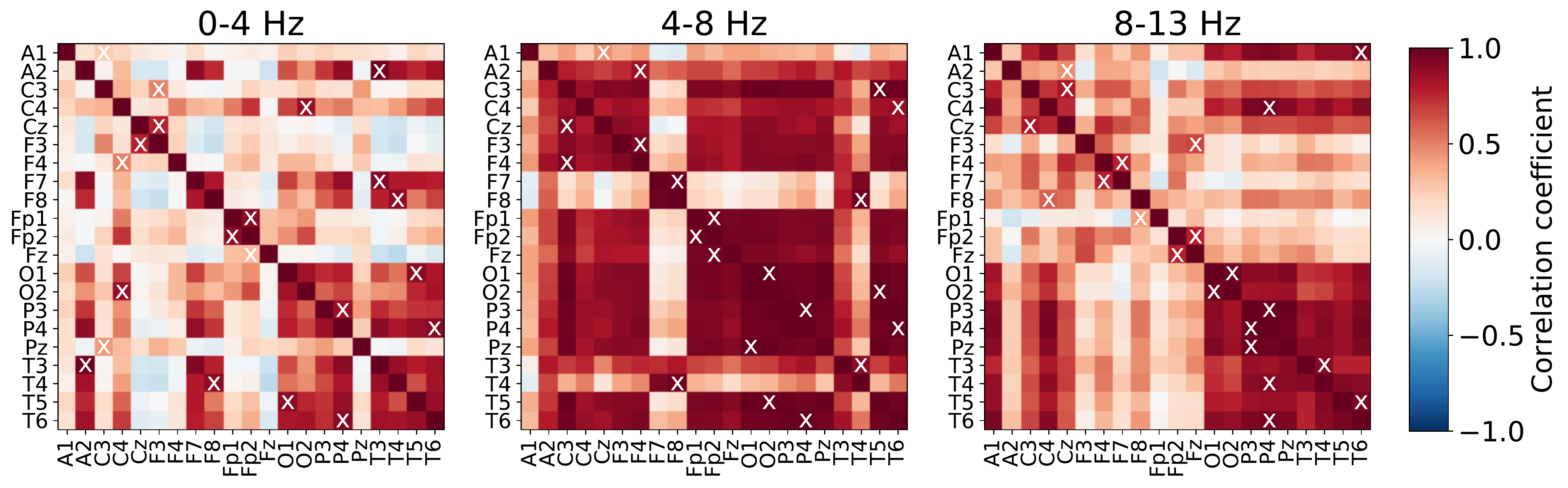}
  \caption{Feature correlation analysis in frequency bands 0--4, 4--8, and 8--13 Hz. For every electrode (row), most highly correlated electrode is marked with a white cross. In frequency band 0--4 Hz, features extracted at T3 are most correlated with those extracted at T4, which could explain the pattern observed in Figure~\ref{fig:important_electrodes}. In the 4--8 Hz frequency region, all features appear to be highly correlated with exceptions such as F7 or F8. In the 8--13 Hz region, features of occipital temporal and parietal region are especially highly correlated.}
  \label{fig:feature_correlations}
\end{figure}

\begin{figure}[ht!]
    \centering
    \includegraphics[width=\textwidth]{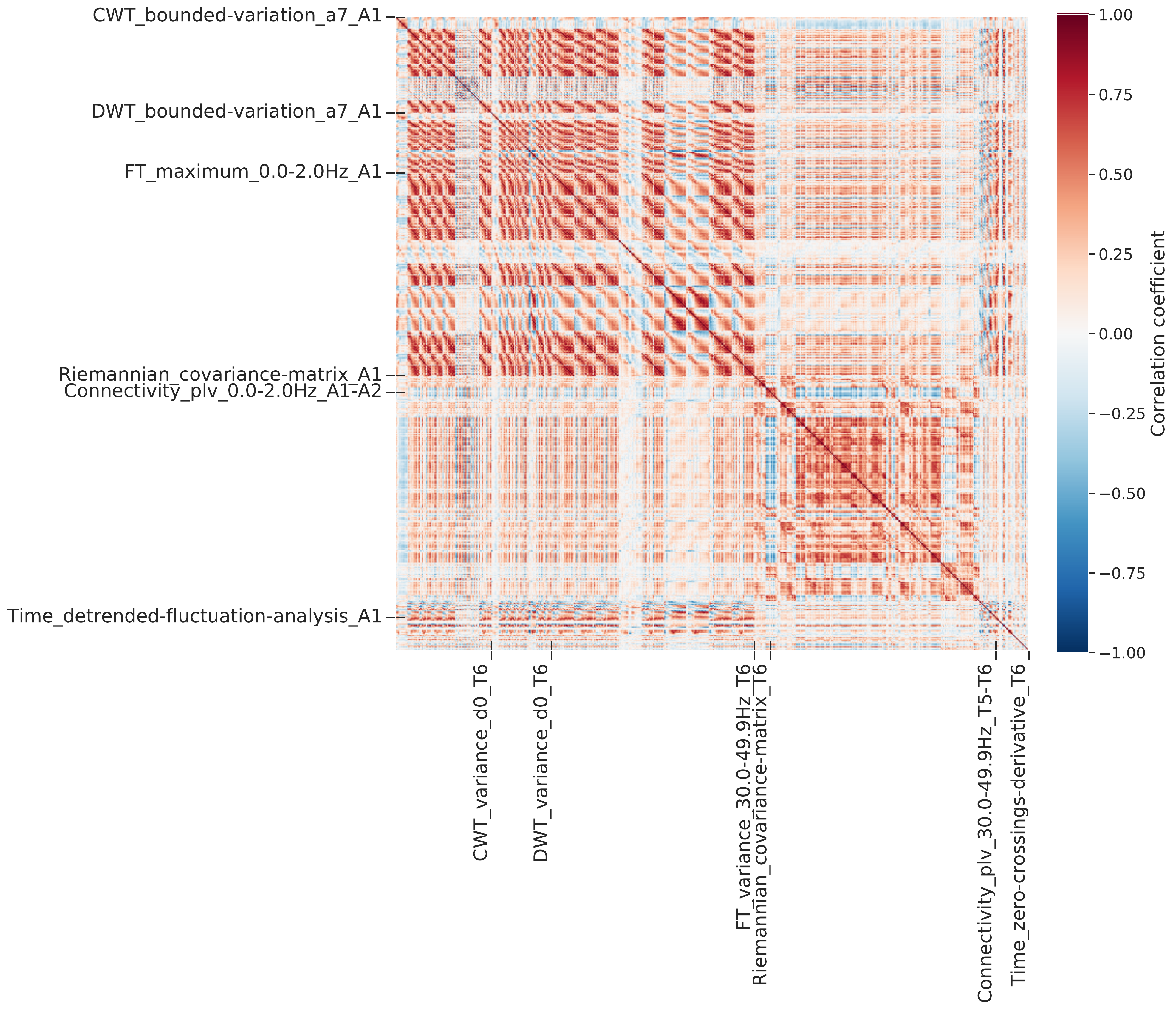}
    \caption{Full feature correlation matrix of all implemented features of domains CWT, DWT, FT, Riemannian, Connectivity, and Time. Features are sorted by feature domain, alphabetically by feature name, ascending by frequency band and alphabetically by electrode location. Strong correlations can be observed across different features within same and between other domains.}
    \label{fig:feature_correlation_matrix}
\end{figure}

To further analyze the features, we computed their correlation coefficients and then visualized these.
We display the full feature correlation matrix [Figure~\ref{fig:feature_correlation_matrix}] and the correlation of FT features extracted from the delta, theta, and alpha range [Figure~\ref{fig:feature_correlations}].
We determined strong correlations, despite feature domain, frequency band, and electrode recording site.

\subsection{Time-resolved feature-based decoding}
\label{sec:result5}
Inspired by the cropped-decoding setup used in our end-to-end pipeline, we implemented a time-resolved (non-aggregated) decoding for our feature-based models. 
However, this did not lead to an increase in decoding accuracy during CV, which is why we did not investigate this setup any further.

Given the drastic increase in the amount of data, it should be possible to realize an improvement using time-resolved decoding, if not here, then for a different task or data set. 
For the given data, the improvements could actually be negligible.
Because the task is concerned with classifying an EEG recording as either pathological or non-pathological, there is likely no information in the signals that evolves at a large time scale.
Because the TUH Abnormal EEG Corpus (v2.0.0) is not a seizure or event data set, we assume that if there exists a continuous alteration in brain activity reflecting a static dysfunction, e.g., related to a structural brain abnormality, it would be indicated consistently. 
This is also one of the key assumptions of our aggregated decoding approach, which performed as well as the ConvNets [see \ref{sec:result1}]. 
If this assumption did not hold, an aggregation, especially using the median as an aggregation function, would result in smoothing of the effect caused by the pathology. 
In a different data set, where signals change over a larger time scale, a time-resolved decoding could result in superior decoding performance. 
However, one must consider the challenges that are posed by time-resolved feature-based decoding. 
In our study, the amount of data increased by a factor of approximately 200 compared to aggregated decoding.

\subsection{Effect of patient-specific information on decoding accuracies}
\label{sec:result6}
We investigated the influence of so-called meta features, i.e., age and gender of patients, with respect to the decoding accuracies. 
Therefore, we added the age and gender of the patients to our feature vectors and as additional network channels to our BD-TCN. 
Classification accuracy improved only marginally during CV, e.g., RF CV accuracy increased by 0.15\% (from 83.1\% to 83.25\%) by adding age and gender. For this reason, we did not further pursue the use of these meta features.
Our result is in consensus with a recent publication, where the authors attempted to combine neural networks and the age information of patients to improve the decoding of pathology from EEGs [\cite{van2019detecting}].
They also found only marginal, insignificant improvements ($+0.07$ area under the receiver operating characteristic curve). 

As clinicians have access to diverse information regarding patients that are not included in the EEG recordings, such as age, gender, current physical and mental condition, patient history, ongoing medical treatment, and more. 
All of this could have a role in their diagnosis or unconsciously introduce a bias. 
For example, we have seen several physician reports in the TUH Abnormal EEG corpus (more than 100) with the comment ``IMPRESSION: Normal EEG for this age'' or similar. 
This is clearly a relative statement by the clinician relying on age information that our models do not have available to find a diagnosis.

Moreover, there is clearly a bias (patient age) in the data set that can be retrieved from the histogram [Figure~\ref{fig:tuh_histogram}]. 
In the data set pathologies certainly appear more frequent with higher age. 
Where this is a justifiable representation of reality, it could unconsciously bias clinicians to diagnose EEG recordings of patients with a high age rather as pathological.
%Apart from the ages and genders of patients, there is even more information hidden in the reports.
Although our experiments did not indicate a substantial improvement when adding patient age and gender in the classification process, the suitable inclusion of patient-specific information from physician reports could improve classification accuracies.

Furthermore, knowledge of other findings related to a patient could unconsciously bias clinicians to be overly sensitive to minimal changes in the EEG.
For example, small changes in an EEG might not be considered as indicative of a pathology on their own, however, in combination with the knowledge of a lesion based on imaging techniques the EEG could be considered as pathological.
Access to different amounts of information could be one explanation for the observation of more false negatives than false positives across all models [Figure~\ref{fig:confusion_matrix}].

\subsection{Ensemble decoding performance}
\label{sec:result7}
\begin{figure}[ht!]
    \centering
\includegraphics[width=0.8\textwidth]{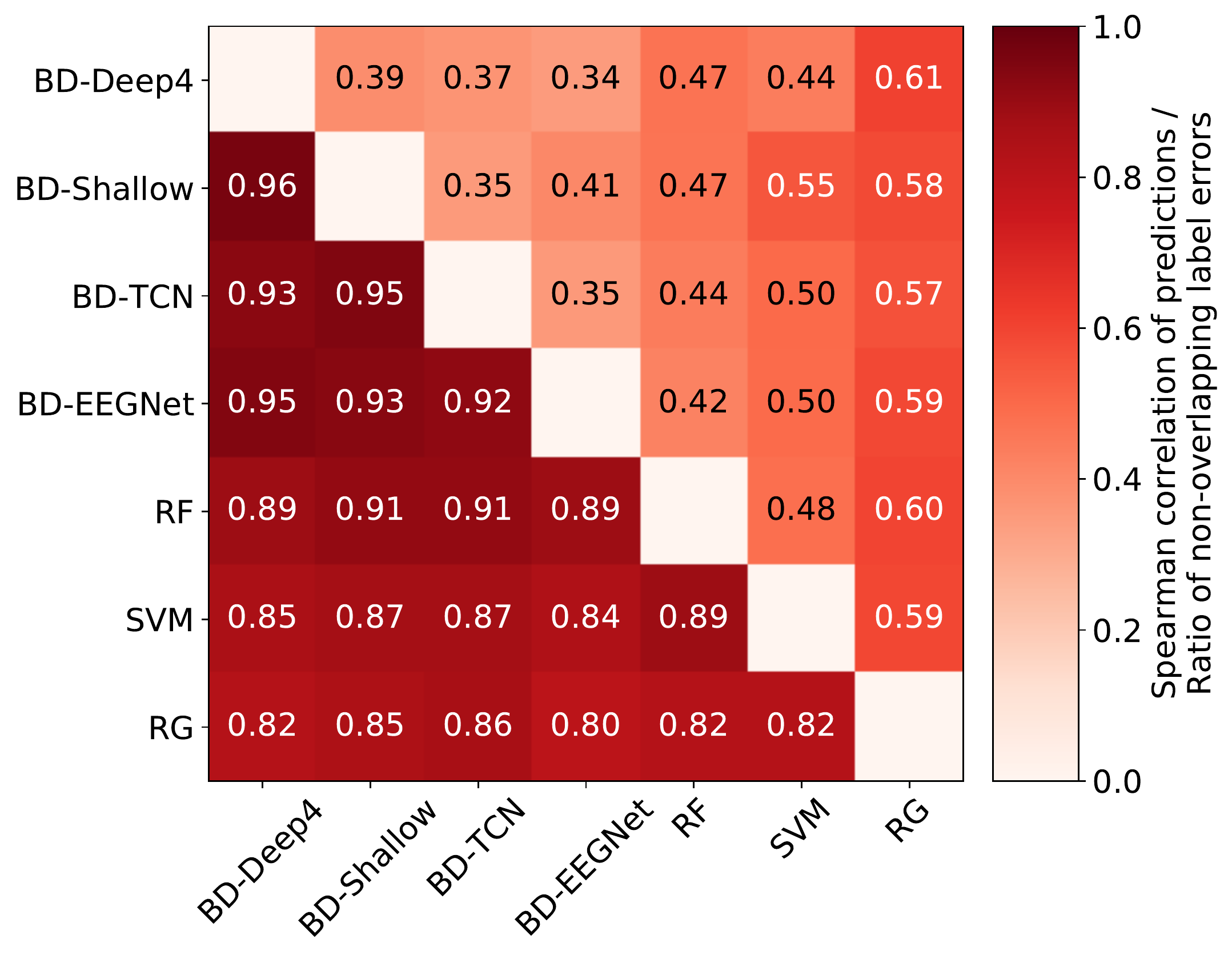}
    \caption{Analysis of CV predictions. In lower triangle, Spearman correlation of pairs of model predictions is displayed. In upper triangle, we added ratio of non-overlapping label errors. Riemannian-geometry-based decoding and BD-Deep4 indicate greatest opportunity for improvement as their predictions have a correlation of 0.82 and ratio of non-overlapping label errors of 0.61.}
    \label{fig:prediction_correlations}
\end{figure}

Our analysis of prediction correlations revealed that the CV predictions of all our models were highly correlated (Spearman correlation of up to 0.96) and indicated a large ratio (up to 61\%) of non-overlapping label errors [Figure~\ref{fig:prediction_correlations}].

\clearpage

We present the results of our two ensemble approaches in Table~\ref{tab:ensembles}.
The combination of the models BD-Deep4, RF, and RG had the highest ratio of non-overlapping CV label errors (336 errors, ratio of 44.56\%), which is why we selected them for majority vote ensembling. 
Automatic ensembling based on auto-sklearn chose every model except the SVM and computed optimal weights based on the CV predictions [Figure~\ref{tab:ensembles}]. 

\begin{table}[ht!]
    \centering
    \begin{tabular}{l|lc|lc}
    & \multicolumn{2}{c|}{\textbf{Majority Vote}} & \multicolumn{2}{c}{\textbf{Auto-sklearn}} \\
    & Model & Weight & Model & Weight \\ \hline
    & BD-Deep4 & 1 & BD-Deep4 & 0.048\\
    & RF & 1 & BD-Shallow & 0.286 \\
    & RG & 1 & BD-TCN & 0.190 \\
    & & & BD-EEGNet & 0.238 \\
    & & & RF & 0.190 \\
    & & & RG & 0.048 \\ \hline
    CV [\%]& \multicolumn{2}{c|}{84.61} & \multicolumn{2}{c}{86.23}\\
    Final evaluation [\%] & \multicolumn{2}{c|}{85.51} & \multicolumn{2}{c}{85.14}\\
    \end{tabular}
    \caption{Result of ensembling investigation indicating models, weights, and performances of two ensembles. One based on majority voting and one automatically selected by auto-sklearn. Neither ensemble improved compared to best single-model performances.}
    \label{tab:ensembles}
\end{table}

Whereas the auto-sklearn ensemble resulted in the overall best CV accuracy (86.23\%), neither ensembling based on majority voting (85.51\%) nor automatic ensemble selection (85.14\%) resulted in an improvement over best single-model performance (BD-TCN: 86.16\%) in the  final evaluation (see Figures~\ref{fig:result_overview}, \ref{fig:confusion_matrix}). 

\section{Discussion}
\label{sec:discussion_new}
In this study, we achieved a feature-based decoding accuracy of 85.87\% using Riemannian-geometry-based decoding [Section~\ref{sec:result1}] which is greater than 7\% higher 
than the previous feature baseline of 78.8\% by \cite{abnormalLopez}. 
Our BD-TCN achieved an even superior decoding accuracy of 86.16\% [Section~\ref{sec:result2}].
We found no statistical evidence that either model under investigation – feature-based or neural network – performed better than the others [Section~\ref{sec:result3}]. 
We observed that temporal EEG dynamics 
in slow frequency ranges were informative for detecting pathological changes, and that this information was used in both pipelines [Section~\ref{sec:result4}].
Moreover, our feature correlation analysis revealed high correlations despite feature domain, frequency band, and electrode recording site.
Neither our time-resolved feature-based decoding
[Section~\ref{sec:result5}], 
nor 
the inclusion of age and gender of patients in the decision processes
resulted in a substantial improvement of decoding accuracy [Section~\ref{sec:result6}]. 
Finally, we determined that ensembling with auto-sklearn can improve classification accuracy in CV [Section~\ref{sec:result7}]; however, not in the final evaluation.

\subsection{Proposed feature-based pathology decoding concept}
\label{disc:feature_approach}
For our study, we aimed at creating a strong feature-based decoding pipeline for a comparison with neural networks.
Therefore, our concept for feature-based decoding involved testing I) a large set of EEG features of different feature domains (in the present study time, amplitude, phase, and connectivity) and II) multiple classifiers including ensembles of classifiers (in the present study RFs, SVMs, and ASC).
In particular, it was important to us to also include recent state-of-the-art feature-based methods such as Riemannian-geometry-based classification [\cite{lotte2018review}].

We found significant differences of our concept to those presented in the literature.
Where we computed 8633 features of 50 feature types and six domains [see Table~\ref{tab:features}], typically far smaller feature sets were used in the literature.
For example, \cite{hosseinifard2013classifying} extracted, exclusively, the total power from the theta, alpha, and beta frequency bands to decode depression.
Similarly, \cite{abnormalLopez} extracted a single feature type (cepstral coefficients) that originated from the field of speech recognition to decode EEG pathology.
An example of a study with a larger amount of feature types is the work of \cite{cai2016pervasive}.
They extracted 16 feature types including amplitude, time, and connectivity measures to detect mild depression.
In our present study, however, we computed an even larger feature set with more than twice the number of feature types [Table~\ref{tab:features}].

In other tasks, such as in the movement of the left hand, right hand, and feet motor decoding applications, it is generally assumed that the primary source of information is contained in the signal amplitudes [\cite{schirrmeister2017hbm}].
Typically, there are a small number of target classes (3--4), which must be modeled. 
For pathology decoding, conversely, the most informative feature domain or type is not \textit{a priori} known.
Furthermore, it is necessary to model a larger number of pathologies.
It cannot be assumed that the pathologies included in the TUH Abnormal EEG Corpus (v2.0.0) can be described by a single feature type.
Therefore, in our concept, we did not \textit{a priori} exclude any feature domains or types.

In our view, the presented concept is not specific to EEG pathology decoding; rather, it is applicable to a wide range of other EEG decoding tasks.
It could well be possible that approaches based on a single feature type, e.g., motor decoding based on amplitude features, could be improved by combining them with different feature types, such as measures of connectivity.

\subsection{Proposed end-to-end pathology decoding concept using deep neural networks}
Similar to the feature-based concept, our aim was also to create a strong end-to-end pipeline.
By end-to-end, we refer to the decoding of raw or minimally preprocessed  data, unlike decoding based on explicitly extracted, handcrafted features.
Recently, end-to-end methods have achieved powerful results in EEG decoding [\cite{craik2019deep}].
In the present study, we applied the two most widely used, well benchmarked, and open source network architectures in EEG decoding: BD-Deep4 [\cite{schirrmeister2017hbm}] and a reimplementation of EEGNet [\cite{lawhern2018eegnet}] as implemented in the Braindecode toolbox.
Furthermore, we also investigated the application of the BD-TCN, an EEG-optimized TCN architecture.
The general TCN architecture was specifically designed to model sequential data and has outperformed RNNs in machine translation and audio synthesis tasks [\cite{bai2018empirical}].
Moreover, we compared to the results of ChronoNet, a CNN/RNN architecture, reported in the literature [\cite{roy2019chrononet}], as it is among the state-of-the-art in EEG pathology decoding.
Finally, we also applied BD-Shallow, which is a special case inasmuch as it was specifically designed to extract band power features.
It is, therefore, between the classical feature-based and end-to-end decoding philosophies.
In summary, the investigated architectures demonstrated significant differences including network archetype (ConvNet, TCN, RNN), parameter count (see Table~\ref{tab:model_parameters}), and number of layers (Figures~\ref{fig:deep}–\ref{fig:shallow}).

\subsection{Feature-based pathology decoding accuracies}
Riemannian-geometry-based decoding resulted in an accuracy of 85.87\%, which was the best feature-based decoding result.
Interestingly, in our study, we observed a narrow range of feature-based decoding accuracies of 81 to 86\% across different classifiers [see Figures~\ref{fig:result_overview}, \ref{fig:confusion_matrix}].
Actually, no statistically significant difference of accuracies of all presented feature-based models could be detected [see Figure~\ref{fig:sign_test}].
Previously, \cite{abnormalLopez} reported an accuracy of 78.8\% based on a single feature type and a CNN/MLP architecture for classification.
Both, the size of the feature sets and choice of the decoders could explain the difference in accuracy.
To the best of our knowledge, the study of \cite{abnormalLopez} is the only previously published feature-based study on EEG pathology decoding.
Compared to this previous feature-based state-of-the-art, our findings demonstrate that substantially higher feature-based decoding accuracies are possible using the proposed approach.

\subsection{Deep end-to-end pathology decoding accuracies}
End-to-end methods yielded accuracies in the range of 82--86\% [see Figures~\ref{fig:result_overview}, \ref{fig:confusion_matrix}].
Astonishingly, this range was extremely similar to the range of accuracy obtained with the feature-based pipeline, which was 81--86\%.
The EEG-optimized TCN was our best-performing neural network architecture achieving an accuracy of 86.16\%.
Interestingly, [\cite{van2019detecting}] also reported an accuracy of approximately 82\% using BD-Deep4 to decode pathology; however, this was based on a different EEG data set.
Their data set is comprised of more examples than the TUH Abnormal EEG Corpus (v2.0.0); however, it is not publicly available.
Similar to the feature-based methods for the architectures under investigation, no statistically significant difference in decoding accuracy could be detected [see Figure~\ref{fig:sign_test}].
Compared to our BD-TCN, ChronoNet achieved a marginally higher decoding accuracy of 86.56\% [\cite{roy2019chrononet}].
In line with our present findings, in a previous study, \cite{heilmeyer2018large} also found no statistically significant difference in accuracies comparing BD-Deep4 with BD-EEGNet in a large-scale benchmark test across different tasks and data sets.
In the following, we now further discuss the virtually identical accuracy range of our feature-based and end-to-end pipelines.

\subsection{Deep end-to-end versus feature based-decoding accuracies}
A main finding of our present study, together with the results of \cite{van2019detecting} and \cite{roy2019chrononet} is that EEG pathology decoding accuracies lie in a narrow range of 81--86\%, even though we compared a broad range of:

\begin{itemize}
    \item \textbf{Analysis strategies} including deep-end-to-end, feature-based, automated ML, and based on RG;
    \item \textbf{Network archetypes} including ConvNet, TCN, and RNN;
    \item \textbf{Network architectures} including BD-Deep4, BD-TCN, BD-EEGNet, BD-Shallow, and ChronoNet;
    \item \textbf{Feature-based classifiers and ensembles} including RF, SVM, and ASC, and;
    \item \textbf{Data sets} including the Temple University Hospital Abnormal EEG Corpus (v2.0.0) and the data set used by \cite{van2019detecting}.
\end{itemize}

Importantly, this range was also considerably below a perfect classification score (100\%).
Decades of previous EEG research have indicated that inter-rater reliability in EEG diagnostics is only moderate [\cite{grant2014eeg}, \cite{houfek1959reliability}, \cite{rose1973reliability}], which ultimately results in label noise.
Regarding label noise, we refer to expert mistakes in diagnosing the EEG recordings as either pathological or non-pathological, which has a number of mutually related consequences [\cite{frenay2013classification}]:
decrease of decoding performance, increase of required amount of data to achieve acceptable decoding performance, and increase of model complexity to properly fit the data.
Furthermore, label noise complicates the identification of relevant features [\cite{frenay2013classification}].
Importantly, in our setting, low inter-rater agreement and the resulting label noise imposed a limit on the theoretically achievable decoding accuracies because we were required to evaluate against the noisy labels within our separate, final evaluation set. Moreover, we did not have access to any rater-independent ground truth\footnote{One can design alternative decoding tasks different from pathology decoding based on the TUH Abnormal EEG Corpus, such as the decoding of patient gender or age.}.
Interestingly, inter-rater agreement in binary classification of EEGs into pathological and non-pathological has been reported as 86--88\% [\cite{houfek1959reliability}, \cite{rose1973reliability}], although these scores were based on EEG ratings of only two neurologists.
Given these numbers, it would appear to be a possibility that EEG pathology decoding accuracies as observed herein and previously by \cite{van2019detecting},  \cite{schirrmeister2017patho}, and \cite{roy2019chrononet} at approximately 86\% could have approached the theoretical optimum imposed by label noise. 
This proposed hypothesis could be tested in the future; however, it would involve a considerable effort.
It would require both a large data set as used in the present study and independent ratings of multiple EEG experts.
A massive amount of EEG data is waiting in the archives of medical centers to be used.
As end-to-end approaches are typically data intense, a larger amount of training data could help to improve generalization further.
Moreover, in traditional feature-based approaches, more data would probably illicit positive effects, as more data in ML is typically favorable over more complex classification algorithms [\cite{halevy2009unreasonable}].
Furthermore, a labeling of EEG recordings as pathological or non-pathological through an ensemble of considerable number of neurologists and/or epileptologists would be a significant beginning to improve label quality.
The inter-rater agreement scores that result from the ensemble could then be included in the data set on a per-recording basis, such that they can be included in a detailed analysis.

In the case where the hypothesis is incorrect and the theoretically optimal EEG pathology decoding accuracy is higher, we see two possible, nonexclusive causes.
The first is that inter-rater reliability of the current data sets is higher than the numbers typically reported in the literature [\cite{houfek1959reliability}, \cite{rose1973reliability}].
The second is that all the methods investigated to date did not extract or use certain features and information that was used by the physicians to determine a diagnosis.
In the first case, the question arises why neither the end-to-end nor the feature-based pipeline could better fit and predict the data.
In the second case, the question arises as to what additional source of information was used by the physicians and how it could be included to enhance performance.
Both cases would open up new, interesting research questions.

\subsection{Learned versus handcrafted features}
Based on our feature visualizations, we determined that features extracted in the theta and delta range from the temporal electrode locations are considered informative.
Knowing that epilepsy is statistically one of the most common disorders of the brain [\cite{thijs2019epilepsy}] and that temporal lobe epilepsy is the most prominent epilepsy [\cite{helmstaedter2009tle}], this could be a reason why this region is important in determining pathology in all decoding pipelines. 
Note that although we know epilepsy is one of the pathologies included in the data set, we neither know the exact number of occurrences nor how many of those suffer from temporal lobe epilepsy.

Interestingly, only features extracted at electrode T4 are considered important in the theta band based on the feature importance analysis [Figure~\ref{fig:important_electrodes}]. 
This is in strong contrast to the network perturbation result [Figure~\ref{fig:perturbation}] and covariance matrix visualization [Figure\ref{fig:covariances_in_ts}], where both hemispheres were considered equally informative in this range.
Given the mechanisms of a decision tree, we assume that the tree chose features extracted at T4 in the theta band early in the decision process, because they are informative. 
Features extracted at T3 in the same frequency band were then not selected for further splitting of examples because they did not provide additional information. 
Hence, we assumed that the features were highly correlated. 

Our feature correlation analysis [Figure~\ref{fig:feature_correlations}] revealed several strong, positive correlations of features extracted at different electrode locations.
Indeed, features extracted at T3 were most correlated with features extracted at T4 (correlation coefficient approximately 0.9) in frequency band 4--8 Hz. 
We actually observed this effect of strong correlations across several of our handcrafted features, despite feature domain, electrode location, and frequency band [Figure~\ref{fig:feature_correlation_matrix}]. 

The covariance matrix visualization [Figure~\ref{fig:covariances_in_ts}] supported the findings of our perturbation and feature importance analysis.
Features extracted from temporal electrode locations were most informative for decoding pathology, especially in the delta and theta frequency range.

The result of the perturbation [Figure~\ref{fig:perturbation}] and feature importance analysis [Figure~\ref{fig:important_electrodes}] in the 8--13 Hz range indicated that occipital electrodes were also extremely informative. 
This is again supported by the covariance matrix visualization [Figure~\ref{fig:covariances_in_ts}]. 
The variance of electrodes O1 and O2 (and also Fp1 and Fp2) are the only variances that were indicative of normal brain activity. 
We explain this by the prominent alpha rhythm in adults in a resting state.
Given the perturbation result, lower amplitudes are anti-correlated with pathology; or, in other words, higher amplitudes are correlated with normal brain activity.

Through the comparison of our analyses, we determined that the first visualization of the feature importance [Figure~\ref{fig:important_electrodes}] was misleading, owing to strong feature correlations. 
Typically, ease of interpretation is considered an advantage of feature-based approaches over end-to-end approaches; however, we determined that there are also limitations to this interpretability.
Pitfalls with RF feature correlations were also described by \cite{altmann2010permutation}.

\subsection{Beyond automated binary EEG pathology diagnostics}
In general, the task we addressed (classifying EEG recordings as either pathological or non-pathological) is especially difficult.
The TUH Abnormal EEG data set does not only contain multiple, highly different pathologies including, but not limited to, Alzheimer's disease, strokes, depression, and epilepsy (which are difficult to model), but also there is only one common label for all of them.
Hence, it can be interpreted as a combination of several decoding tasks.
Presumably, every individual decoding task has a non-convex loss landscape.
The performance of deep neural networks heavily depends on the optimization of the loss function and hence finding a minimum in the loss landscape [\cite{li2018visualizing}].
A landscape that arises from the combination of tasks could be even more complex and, therefore, be more difficult to optimize.

To investigate whether the decoding of individual pathologies is easier and offers a path towards automatic EEG diagnostics beyond binary classification, we initiated work on restructuring the physician reports included in the TUH Abnormal EEG Corpus to a tabular format which is a more machine and user-friendly representation.
%and created a spreadsheet\footnote{Spreadsheet is available for download and collaboration at \url{https://github.com/gemeinl/auto-eeg-diagnosis-comparison}}, where we attempt to represent the reports in a more machine and user-friendly manner. 
%Through the sharing of this version of the reports, 
The code for restructuring is uploaded to our repository\footnote{available for download at \url{https://github.com/gemeinl/auto-eeg-diagnosis-comparison}}.
We are seeking support to further specify the diagnoses of patients (to ICD-10 codes [\cite{world1992icd}]) whose recordings were labeled pathological.
Moreover, we aim to find a suitable method to encode patient histories and ongoing medical treatment, such that they can be used in classification. 

Although we are extremely interested in further specifying the diagnoses of patients included in the physician reports as a basis for advanced pathology decoding, we doubt its clinical usefulness. 
Owing to their ease of application, low cost, wide availability, and non-invasive nature, EEGs are a valuable screening method. 
Typically, when an EEG recording is found to be pathological, multiple other clinical sources, i.e., medical imaging such as magnetic resonance imaging (MRI), are involved to actually determine a diagnosis. 
In the majority of cases, an EEG is not used (and not sufficient) to diagnose a neurological disease or disorder, as an occurrence of pathological change is seldom specific for one disease or disorder. 
From our point of view, a reasonable extension of the given decoding task could be to classify recordings as either normal, epileptic, or differently pathological. 
As epilepsy is one of the most common disorders of the brain [\cite{thijs2019epilepsy}], this classification would further support neurologists in screening patients using EEGs.

\subsection{Clinical usefulness of current decoders of EEG pathology}
Everyday medical applications typically require higher accuracy than the current state-of-the-art in EEG pathology decoding to be accepted.
However, decoding pipelines with an accuracy in the current range can be valuable.
For example, they could make EEG diagnostics available to patients that cannot attend specialized centers.
This includes wide areas of developing countries where specialized centers and neurological experts are rare. 
Approximately 50 million people worldwide suffer from epilepsy [\cite{world2019epilepsy}], of which the vast majority live in developing countries [\cite{world2019epilepsy}, \cite{singh2016epidemiology}].
In these countries, in addition to their disease or disorder, patients frequently suffer from social stigma [\cite{world2019epilepsy}, \cite{newton2012epilepsy}] owing to missing diagnoses and inexperience in addressing those diseases and disorders.
In our opinion, although diagnostic accuracy is at approximately 86\%, an automatic diagnosis is better than not being diagnosed at all.
The pipelines could be used as a prescreening method, which could recommend a visit to a specialized center in the case of the detection of pathological activity.

\subsection{Implications for EEG decoding accuracy evaluation in general}
Our findings of statistically similar decoding accuracies of different networks and classifiers has implications for other publications.
EEG decoding papers based on deep end-to-end learning frequently compare their results to only (rather) simple feature-based approaches or exclusively to other deep end-to-end learning results.
For example, for the pathology decoding task based on the TUH Abnormal EEG corpus, all publications based on deep end-to-end learning compared to the result of \cite{abnormalLopez}.
This leads to the impression that deep neural networks heavily outperform feature-based approaches.
However, in this study, we demonstrated that with a somewhat elaborate feature-based approach, one can achieve decoding results similar to deep end-to-end methods [Figures~\ref{fig:result_overview}, \ref{fig:confusion_matrix}].
To be more precise, there is actually no statistical evidence that the investigated networks perform any better than the feature-based approaches [Figure~\ref{fig:sign_test}].
We therefore emphasize once more, that fair comparisons to strong baselines are essential to assess the quality of decoding results.
This does not only hold for EEG pathology decoding tasks, but also for EEG decoding tasks in general.
Moreover, the necessity of comprehensive comparisons holds for all areas of application of ML.

\subsection{Public availability of resources and reproducibility}
The availability of code that can be used to independently reproduce the published results is currently the only method to truly assess the scientific quality and generalizability of the proposed approaches. 
To the best of our knowledge, there are only six other published results for general pathology decoding from EEGs, five of which are based on the TUH Abnormal EEG Corpus (Table~\ref{tab:related_works}). 
\cite{van2019detecting} have used and adapted our BD-Deep4 network and applied both the original and adapted version to the same task on a different, even larger data set.
Their results indicate that both versions yield identical decoding results in a similar range (82\% accuracy) to what we present in this study.
Furthermore, they also investigated the effect of including patient age in the classification process and determined only marginal improvements, which is in consensus with our observations [see \nameref{sec:result6}].
First, this underlines the general applicability of our BD-Deep4; second, it serves as a reproduction study of our results.
To improve the reproducibility in EEG decoding further, we have uploaded the resources\footnote{\url{https://github.com/robintibor/auto-eeg-diagnosis-example}} of our previous study [\cite{schirrmeister2017patho}] and also uploaded the resources of our current study\footnote{\url{https://github.com/gemeinl/auto-eeg-diagnosis-comparison}}. 

\subsection{Potential improvements of the decoding pipelines}
Although we implemented a large set of features of different domains for this study, this collection is not close to completion.
There are probably an infinite number of features one could implement, e.g., in the domain of connectivity features one could additionally investigate the usage of cross-correlation, cross-coherence, mutual information, omega complexity, s-estimator, and global field synchronization [\cite{jalili2013synchronization}].
We have already attempted to add the gender and ages of the patients to improve classification [\ref{sec:result6}].
In clinical diagnostics, physicians have access to even more information such as medical patient history and ongoing medication.
It is an open challenge as to how to include this information in the decision process.

With extremely high-dimensional feature-spaces as presented in this study, it is natural to rethink dimensionality reduction methods.
A small feature dimension is favorable because it yields shorter learning times and makes interpretation easier.
Although our preliminary feature selection with PCA resulted in a decrease of decoding performance, there are several other approaches to be investigated, including independent component analysis [\cite{comon1994independent}] (ICA) or tensor decomposition [\cite{sidiropoulos2017tensor}].
Furthermore, one could also attempt to enhance signal quality by applying a source reconstruction method [\cite{michel2004eeg}] prior to the extraction of the features and classification.
We have observed that features extracted from temporal electrode locations were highly informative for the decoding of EEG pathology [\ref{sec:result4}.
A more precise localization of signals could also improve classification accuracy.

As in the implementation of features, there are numerous other feature-based classifiers available.
ASC automatically selected models that we did not choose ourselves [Section~\ref{sec:result7}], i.e., AdaBoost and Gradient Boosting.
Both classifiers, however, are frequently implemented with decision trees that also form the basis of RFs.
It could be worth further investigation into the application of other classifiers.
A specific case is when performing time-resolved feature-based decoding [Section~\ref{sec:result5}], where the models chosen for the present study could not yet benefit from the drastic increase of data.

Finally, one could run a pipeline search optimization based on automated ML [\cite{automl_book}].
This could not only include feature scaling, selection of classifiers, and optimization of their hyperparameters as executed by ASC and as already performed for the present study.
One could also attempt to optimize the hyperparameters for feature extraction itself, such as frequency bands, aggregation function, window length, window function, and others.
We are convinced that a systematic optimization of these choices could lead to even better-performing feature-based EEG decoding results.

In the huge search space of network architectures, the models investigated in this study possibly lie in a subspace that contains well-performing architectures for this task.
Most likely, these architectures are not yet optimal to decode EEG pathology.
Many research groups are now promoting the development of neural architecture search (NAS) [\cite{elsken19_nas}] to address the concern that handcrafting features is now being replaced by architecture crafting.
NAS is achieving rapid progress and we assume that it will spawn even better-performing architectures in the close future.

\section{Outlook and Conclusion}
\label{sec:outlook}
We emphasize once more that the discussed study with a large data set and improved inter-rater quantification should be performed to test our hypothesis of an accuracy limit in EEG pathology decoding at approximately 90\%. 
In addition to the possibilities for improvements discussed above, data augmentation could be another.
Although augmentation for EEGs is not as intuitive as for other types of signals, the first promising results used cutout on raw EEG signals [\cite{pereira2018cross}] or rotation of electrode recording sites [\cite{krell2018data}]. 
Augmentation could also be combined with automated ML [\cite{cubuk2018autoaugment}, \cite{lim2019fast}] to further advance the data intense end-to-end models.

To circumvent the consequences of label noise in EEG pathology decoding, instead we propose to decode I) the gender of patients to better assess the potentials of feature-based and end-to-end pipelines,
and II) the age of patients to use the gap of chronological age to predicted brain age as an alternative source for indication of pathology.
In the literature this gap and its estimation based on MRI scans is referred to brainAGE [\cite{franke2013brainage}, \cite{franke2010estimating}].

Based on our insights obtained through the present study, we see a promising future for automated EEG diagnostics.
A well-working pipeline that implements the mentioned options for improvements could be helpful in the interpretation of EEG recordings.
It could not only make EEG diagnostics available to patients that cannot attend specialized centers, but could also allow for earlier detection of pathologies on the same level as ensembles of human experts, and thereby somewhat reduce the global burden of neurological diseases and disorders.

\section*{Acknowledgements}
This work was supported by the Graduate School of Robotics, Georges-K\"ohler-Allee 80, 79110, Freiburg, Germany and the State Graduate Funding Program of Baden-W\"urttemberg, Germany.

\section*{Conflict of interest}
The authors declare no competing financial interests.

\section*{Data and code availability statements}
The TUH Abnormal EEG Corpus [\cite{abnormalLopez}] used for our study is a subset of the TUH EEG Corpus [\cite{obeid2016temple}, DOI: 10.3389/fnins.2016.00196].
Both data sets are publicly available for download upon registration at \url{www.isip.piconepress.com/projects/tuh\_eeg/html/downloads.shtml}. 

The code used for our study relies on the open source toolboxes Braindecode (\url{github.com/TNTLFreiburg/braindecode}) and Brainfeatures (\url{github.com/TNTLFreiburg/brainfeatures}). 
Code specific to the experiments performed for our study was uploaded to \url{github.com/gemeinl/auto-eeg-diagnosis-comparison}.

%\bibliographystyle{apalike}
%\bibliography{bibliography}

\appendix
\beginsupplement

\section{Hyperparameters}
In this section, we report all the hyperparameters used for all the investigated models in our experiments to allow maximal reproducibility.

\begin{table}[ht!]
    \centering
    \begin{tabular}{c|c|c|c|c|c}
    Hyperparameter & RF & SVM & Riemannian (SVM) & SVM (on covariance matrices) & ASC \\ \hline
    C & & 100 & 10 & 1000 & \\
    gamma &  & $ 1 / \sqrt{F} $ & $ 1 / (E * (E + 1) / 2) $ & $ 1 / (E * (E + 1) / 2) $ &\\ \hline
    criterion & entropy & & & & \\
    bootstrap & False & & & & \\
    max\_features & $ \sqrt{F} $ &  & & & \\
    min\_sampels\_leaf & 2 & & & & \\
    min\_samples\_split & 2 & & & & \\
    max\_depth & 90 & & & & \\
    n\_estimators & 1600 & & & & \\ \hline
    Total run time limit & & & & & 12h\\
    Per run time limit & & & & & 4h\\
    \end{tabular}
    \caption{Hyperparameters of feature-based approaches. Optimized using CV.}
    \label{tab:feature_hyperparameters}
\end{table}

\begin{table}[ht!]
    \centering
    \begin{tabular}{c|c|c|c|c}
    Hyperparameter & BD-Deep4 & BD-Shallow & BD-TCN & BD-EEGNet \\ \hline
    batch\_size & 64 & 64 & 64 & 64 \\
    max\_epochs & 35 & 35 & 35 & 35 \\
    n\_start\_chans & 25 & 40 & & \\
    n\_chans & 21 & 21 & 21 & 21 \\
    final\_conv\_length & 1 & 25 & & 18 \\
    n\_chan\_factor & 2 & & & \\
    model\_constraint & None & None & None & None \\
    init\_lr & 0.01 & 0.000625 & 0.0011261049710243193 & 0.001 \\
    weight\_decay & 0.0005 & 0 & 5.83730537673086e-07 & 0 \\
    l2 decay & & & 1.7491630095065614e-08 & \\
    input\_time\_length & 6000 & 6000 & 6000 & 6000 \\
    stride\_before\_pool & True & & & \\
    dropout & 0.5 & 0.5 & 0.05270154233150525 & 0.25 \\
    \end{tabular}
    \caption{Hyperparameters of neural netowkrs. Optimized in previous experiments (default values were used for BD-EEGNet).}
    \label{tab:network_hyperparameters}
\end{table}

\section{Performance differences to older experiments}
We determined marginally different results for our BD-Deep4 and BD-Shallow decoding pathology compared to older experiments. In our previous publication \cite{schirrmeister2017patho}, we reported a final decoding accuracy of 85.4\% using BD-Deep4 (here 84.57\%) and 84.5\% using BD-Shallow (here 84.13\%). We assume that the differences are due to the other optimizations used in training the networks. Where we previously used max-norm constraint, no weight decay, optimizer Adam, and a fixed learning rate, we now used cosine annealing to schedule learning rate and weight decay, and used optimizer AdamW. In certain experiments, we observed updates in the last epochs that decreased the classification accuracy. It could be possible that with this updated optimization, other hyperparameters must be adjusted to yield even more robust and superior results. Furthermore, the data set has been updated. The current version is TUH Abnormal EEG Corpus (v2.0.0); in our previous publication, we used version (v1.1.1). The updates included removing duplicates that were present in both training and final evaluation sets, which possibly made the classifications of these recordings easier.

\section{Neural network parameters}
In Table~\ref{tab:model_parameters} we display the number of parameters of each investigated neural network architecture. 
They were computed by summation of the sizes of layers as returned by PyTorch\footnote{available for download at \url{https://pytorch.org/get-started/locally/}} [\cite{paszke2017automatic}].
The complexity of models from low to high is BD-EEGNet, BD-Shallow, BD-Deep4,and BD-TCN. 
In our analysis, models with more parameters achieved higher accuracies.

\begin{table}[ht!]
    \centering
    \begin{tabular}{c|c|c}
        Model & Parameters $ [\times 10^3] $ & Receptive field \\ \hline
        BD-TCN & 457 & 903 \\
        BD-Deep4 & 277 & 600 \\
        BD-Shallow & 37 & 608 \\
        BD-EEGNet & 7 & 597 \\
    \end{tabular}
    \caption{Investigated neural network architectures and related number of parameters and receptive field sizes. BD-TCN is most complex and has largest receptive field, BD-EEGNet is, significantly, the simplest model.}
    \label{tab:model_parameters}
\end{table}

\section{Riemannian-geometry-based decoding}
We investigated the influence of averaging covariance matrices for Riemannian-geometry-based decoding using Euclidean and geometric mean, and tangent space mapping [Figure~\ref{tab:riemann_comparison}].
\begin{table}[ht!]
    \centering
    \begin{tabular}{c|c|c|c}
        \textbf{Mean} & \textbf{Subset} & \textbf{Tangent Space} & \textbf{Accuracy [\%]} \\ \hline
        Euclidean & Cross-Validation & False & 73.64 \\
         & & True & 80.00 \\
         & Final evaluation & False & 76.81 \\
         & & True & 84.42 \\ \hline
        Geometric & Cross-validation & False & 78.57 \\
         & & True & 81.26 \\
         & Final evaluation & False & 81.52 \\
        & & True & 85.87
    \end{tabular}
    \caption{Comparison of Riemannian-geometry-based decoding using arithmetic mean, geometric mean, and optional tangent space mapping. Using tangent space mapping and geometric mean estimation yielded best performance.}
    \label{tab:riemann_comparison}
\end{table}
As could be expected [\cite{barachant2013classification}], aggregation with the geometric mean yields superior decoding accuracies.
Additionally, tangent space mapping yielded substantial improvements.

\section{Three model ensembles errors}
In Table~\ref{tab:ensemble_of_three_errors}, we display the analysis of the errors of the three-tuples of the models for ensembling. We chose those three models (BD-Deep4, RF, RG) that indicated the highest ratio (44,56\%) of CV label errors made by exactly one of the models, as these could be reduced by majority voting.

\begin{table}[ht!]
    \centering
    \begin{tabular}{llll|rrrr|r}
    {} &  Model 1 &  Model 2 & Model 3 &    E0 &   E1 &   E2 &   E3 &  E1 Ratio [\%] \\
    \midrule
    0  &     BD-Deep4&  BD-Shallow&     BD-TCN &  2163 &  166 &  110 &  277 &         30.02 \\
    1  &     BD-Deep4&  BD-Shallow&  BD-EEGNet&  2130 &  177 &  123 &  286 &         30.20 \\
    2  &     BD-Deep4&  BD-Shallow&      RF &  2084 &  240 &  127 &  265 &         37.97 \\
    3  &     BD-Deep4&  BD-Shallow&     SVM &  2053 &  255 &  149 &  259 &         38.46 \\
    4  &     BD-Deep4&  BD-Shallow&      RG &  2006 &  299 &  193 &  218 &         42.11 \\
    5  &     BD-Deep4&      BD-TCN &  BD-EEGNet&  2158 &  150 &  121 &  287 &         26.88 \\
    6  &     BD-Deep4&      BD-TCN &      RF &  2108 &  218 &  127 &  263 &         35.86 \\
    7  &     BD-Deep4&      BD-TCN &     SVM &  2084 &  227 &  140 &  265 &         35.92 \\
    8  &     BD-Deep4&      BD-TCN &      RG &  2021 &  304 &  166 &  225 &         43.74 \\
    9  &     BD-Deep4&   BD-EEGNet&      RF &  2092 &  201 &  145 &  278 &         32.21 \\
    10 &     BD-Deep4&   BD-EEGNet&     SVM &  2056 &  238 &  138 &  284 &         36.06 \\
    11 &     BD-Deep4&   BD-EEGNet&      RG &  1990 &  312 &  179 &  235 &         42.98 \\
    12 &     BD-Deep4&       RF &     SVM &  2039 &  241 &  175 &  261 &         35.60 \\
    \textbf{13} &     \textbf{BD-Deep4} &       \textbf{RF} &      \textbf{RG} &  \textbf{1962} &  \textbf{336} &  \textbf{207} &  \textbf{211} &         \textbf{44.56} \\
    14 &     BD-Deep4&      SVM &      RG &  1953 &  322 &  221 &  220 &         42.20 \\
    15 &  BD-Shallow&      BD-TCN &  BD-EEGNet&  2144 &  178 &  107 &  287 &         31.12 \\
    16 &  BD-Shallow&      BD-TCN &      RF &  2111 &  216 &  122 &  267 &         35.70 \\
    17 &  BD-Shallow&      BD-TCN &     SVM &  2056 &  272 &  134 &  254 &         41.21 \\
    18 &  BD-Shallow&      BD-TCN &      RG &  2033 &  286 &  166 &  231 &         41.87 \\
    19 &  BD-Shallow&   BD-EEGNet&      RF &  2077 &  227 &  138 &  274 &         35.52 \\
    20 &  BD-Shallow&   BD-EEGNet&     SVM &  2024 &  269 &  172 &  251 &         38.87 \\
    21 &  BD-Shallow&   BD-EEGNet&      RG &  1993 &  295 &  204 &  224 &         40.80 \\
    22 &  BD-Shallow&       RF &     SVM &  2012 &  278 &  182 &  244 &         39.49 \\
    23 &  BD-Shallow&       RF &      RG &  1970 &  325 &  205 &  216 &         43.57 \\
    24 &  BD-Shallow&      SVM &      RG &  1935 &  343 &  233 &  205 &         43.92 \\
    25 &      BD-TCN &   BD-EEGNet&      RF &  2108 &  195 &  137 &  276 &         32.07 \\
    26 &      BD-TCN &   BD-EEGNet&     SVM &  2059 &  240 &  153 &  264 &         36.53 \\
    27 &      BD-TCN &   BD-EEGNet&      RG &  2010 &  305 &  161 &  240 &         43.20 \\
    28 &      BD-TCN &       RF &     SVM &  2049 &  236 &  183 &  248 &         35.38 \\
    29 &      BD-TCN &       RF &      RG &  1992 &  313 &  191 &  220 &         43.23 \\
    30 &      BD-TCN &      SVM &      RG &  1962 &  331 &  204 &  219 &         43.90 \\
    31 &   BD-EEGNet&       RF &     SVM &  2018 &  253 &  178 &  267 &         36.25 \\
    32 &   BD-EEGNet&       RF &      RG &  1960 &  321 &  207 &  228 &         42.46 \\
    33 &   BD-EEGNet&      SVM &      RG &  1925 &  346 &  221 &  224 &         43.74 \\
    34 &       RF &      SVM &      RG &  1924 &  337 &  234 &  221 &         42.55 \\
    \end{tabular}
    \caption{Overview of all investigated ensembles of three models indicating  number of label errors made by none of the models (E0), by exactly one (E1), by two (E2) or by all of the models (E3) and indicating ratio of E1 errors that could potentially be reduced by combining three models in an ensemble. Highlighted models were chosen for ensembling.}
    \label{tab:ensemble_of_three_errors}
\end{table}

\end{document}